\documentclass[useAMS,usenatbib]{mn2e}
\usepackage{times}
\usepackage{graphicx}
\usepackage{epstopdf}
\newcommand{\mdtyEG}{eg.} 

\newcommand{\eqn}[1]{equation~(\ref{#1})}
\newcommand{\eqns}[1]{equations~(\ref{#1})}
\newcommand{\Eqn}[1]{Equation~(\ref{#1})}
\newcommand{\Eqns}[1]{Equations~(\ref{#1})}
\newcommand{\eqNo}[1]{(\ref{#1})}
\newcommand{\figref}[1]{Fig.~\ref{#1}}
\newcommand{\Figref}[1]{Fig.~\ref{#1}}
\newcommand{\otherfigref}[1]{fig.~{#1}}
\newcommand{\tableref}[1]{Table~\ref{#1}}
\newcommand{\tablerefs}[1]{Tables~\ref{#1}}
\newcommand{\Tableref}[1]{Table~\ref{#1}}
\newcommand{\othertableref}[1]{table~{#1}}
\newcommand{\othertablerefs}[1]{tables~{#1}}
\newcommand{\secref}[1]{Section~\ref{#1}}
\newcommand{\Secref}[1]{Section~\ref{#1}}

\newcommand{\nAPP}[1]{\ensuremath{n_{\mathrm{#1}}}}

\newcommand{\SpinDownAge}{\ensuremath{t_{s}}}

\newcommand{\LRdefined}{\ensuremath{\equiv}}

\title[Pulsar magnetic alignment and the pulsewidth-age
relation]{Pulsar magnetic alignment and the pulsewidth-age relation}
\author[M. D. T. Young, L. S. Chan, R. R. Burman and D. G. Blair]
{M. D. T. Young$^{1}$\thanks{E-mail:~~Matthew.Young@icrar.org}, L. S. Chan$^{2}$, R. R. Burman$^{2}$ and D. G. Blair$^{2}$\\
$^{1}$International Centre for Radio Astronomy Research, M468, University of Western Australia, 7 Fairway, Crawley, WA 6009, Australia\\
$^{2}$School of Physics, M013, University of Western Australia, 35 Stirling Hwy, Crawley, WA 6009, Australia}
\begin{document}
\date{Accepted 2009 October 30. Received 2009 October 30; in original form 2009 October 07}
\pagerange{\pageref{firstpage}--\pageref{lastpage}} \pubyear{2009}
\maketitle
\label{firstpage}
\begin{abstract}
    Using pulsewidth data for 872 isolated radio pulsars we test the 
    hypothesis that pulsars evolve through a progressive narrowing of
    the emission cone combined with progressive alignment of the spin and 
    magnetic axes.  The new data provide strong evidence
    for the alignment over a time-scale of about $1\,$Myr
    with a log standard deviation of around 0.8 across the observed
    population.  This time-scale is shorter than the time-scale of
    about $10\,$Myr found by previous authors, but the log standard
    deviation is larger.  The results are inconsistent with models
    based on magnetic field decay alone or monotonic counter-alignment to
    orthogonal rotation. The best fits are obtained for a braking index parameter, 
    $n_{\gamma}\approx2.3$, 
    consistent the mean of the six measured values, but based on a much larger sample of 
    young pulsars.    
    The least-squares fitted models are used to predict the
    mean inclination angle
    between the spin and magnetic axes as a function of log characteristic age. 
    Comparing these predictions 
    to existing estimates it is found that the model
    in which pulsars are born with a random angle of inclination gives
    the best fit to the data.   Plots of the mean beaming fraction as a
    function of characteristic age are presented using the
    best-fitting model parameters.  
\end{abstract}

\section{INTRODUCTION}

One of the greatest of the many mysteries of radio pulsars is their
evolutionary histories.  As pulsars age, powerful electromagnetic torques
act to increase the rotation period, $P$, of the strongly magnetized
neutron star, causing the spin-frequency, $\nu=1/P$, to decrease over
time, a phenomenon known as spin-down.  This can be described by the
equation
\begin{equation}
    \dot{\nu}= -K \nu^{n}
    \mbox{~,}
    \label{eq:SpinDown}
\end{equation}
where $n$ is a parameter known as the braking index, with $n=3$ for a
magnetic dipole rotating in a vacuum.  The spin-down torque acting on
the star is proportional to $\dot{\nu}$, and it decays over time as
$\dot{\nu}$ decreases.

There is also evidence that $K$ is not constant, but rather is also changing over time.  
The surface dipole magnetic field strength at the magnetic equator, assuming that the 
spin and magnetic axes are orthogonal, is conventionally given by  
$B_{\mathrm{surf}}=3.2\times10^{19}(P\dot{P})^{1/2}\,\mbox{G}$; at the magnetic poles 
the field strength is $2 B_{\mathrm{surf}}$ \citep{ls06}. $B_{\mathrm{surf}}$ tends to 
be bigger for younger pulsars than for older ones having a large spin-down age, 
\(\SpinDownAge{}\LRdefined{}P/\left[(n-1)\dot{P}\right]\).  This in turn suggests that 
$K$ tends to reduce with increasing age, and much conjecture remains about the 
origin(s) of this evolution.

Many authors have argued that the main contributor to the $K$-evolution
is the progressive alignment of the spin and magnetic axes: e.g.
\citet{cb83}, \citet{cb86}, \citet{lm88}, \citet{xw91}, \citet{kw92a},
\citet{can93}, \citet{pp96}, \citet{tm98} and \citet{wj08}.  But \citet{mck93} and
\citet{gh96} have argued in favour of a random distribution of the
inclination angles between the angular velocity and magnetic axes and
against magnetic alignment.  Others have favoured magnetic field decay
\citep[e.g.][]{no90}, while there are also authors who argue for
continuing counter-alignment to eventual orthogonality of the axes,
via an electromagnetic torque exerted by magnetospheric electric
currents flowing on open field lines \citep{bgi84}.

According to \citet{jon76}, pulsars initially counter-align (via a strongly 
temperature-dependent dissipative torque in the fluid interior)
to reach the orthogonal rotator state after about $10^{3}\,$yr, where they remain 
for $10^4$--$10^{5}\,$yr (while the dissipative torque decays as the interior 
cools), after which alignment (via electromagnetic torque) occurs.    

If $K$ were constant, the braking index $n$ would be equal to the
\emph{apparent} braking index,
\begin{equation}
    \nAPP{app} 
    \LRdefined {\nu\ddot{\nu}}/{\dot{\nu}^{2}} 
    = 2 - {P \ddot{P}}/{\dot{P}^{2}}
    \mbox{~.}
    \label{eq:BrakingIndex:Definition}
\end{equation}
If $\ddot{\nu}$ can be determined from observations of a pulsar, then
$\nAPP{app}$ can be computed.  However, if $K$ is
varying, then in general $\nAPP{app}\neq{}n$.  Some authors
\citep[\mdtyEG{}][]{jg99, tk01} have argued that, for pulsars of moderate age,
$\nAPP{app}$ exceeds the value of 3 that corresponds to magnetic
dipole radiation.  This could be taken to imply that either magnetic
alignment or field decay must be occurring.  However, it is likely
that recovery from unseen pulsar glitches is the principal
contribution to measured variations in $\dot{P}$ \citep{wmz+01} for 
these pulsars.

In this paper, we shall use the pulsar evolution models from
\citet[][hereafter CB83]{cb83}, \citet[][hereafter CB86]{cb86} and
\citet[][hereafter C93]{can93}, which all invoke a progressive
alignment of the spin and magnetic axes.  The CB83 model develops the
pulsewidth-age relation on the assumption that all pulsars move along
the same evolutionary track, differing only in age and orientation
with respect to Earth.  The CB86 model relaxes this assumption and
allows a distribution of alignment time-scales, as well as a
time-varying relationship between characteristic and actual age.  The
C93 models additionally allow a distribution of initial inclination
angles.  All of these models predicted a minimum in the mean
pulsewidth as a function of characteristic age and provided a good fit
to the available data, consisting of 293 pulsewidth values from the
\citet{mt81} catalogue.

Here we use the ATNF (Australia Telescope National Facility) pulsar catalogue\footnote{http://www.atnf.csiro.au/research/pulsar/psrcat/} \citep{mhth05}, version 1.35, to investigate pulsar evolution through magnetic alignment using the Candy--Blair models.  This contemporary catalogue contains much larger data sets of 872 and 1420 isolated radio pulsars with 10~per~cent and 50~per~cent intensity pulsewidth values, $W_{10}$ and $W_{50}$.  It also contains many more old pulsars than was previously the case, which tend to have lower radio luminosities.  It thus allows a more precise analysis of pulsar evolution.  As a consistency check, we also examine separately those pulsars with pulsewidths measured during the Parkes multibeam pulsar survey \citep{mlc+01,lfl+06}, giving $W_{10}$ and $W_{50}$ data sets of 377 and 934 pulsars respectively.

The paper is structured as follows.
\Secref{sec:MagneticAlignmentTheory} presents a summary of the theory
of alignment between the spin and magnetic axes;
\secref{sec:PulsewidthEvolution} gives a mathematical description of
pulsewidth evolution; in \secref{sec:Data:Fitting} we least-squares
fit the evolutionary pulsewidth models to the $W_{10}$ and $W_{50}$
data as functions of characteristic age; in \secref{sec:analysis} we
further analyse these results, in particular comparing them to the
inclination angle estimates of \citet{ran93b} and \citet{gou94}; and conclusions
are presented in \secref{sec:Conclusions}.

\section{Magnetic alignment theory}
\label{sec:MagneticAlignmentTheory}

\subsection{Evolution in true age}

The Candy--Blair pulsar evolution models are based on two effects. One is 
progressive alignment of the spin and magnetic axes, for which the formula 
describing the alignment phase of the \citet{jon76} model is adopted: this has an 
exponential decay of the sine of the magnetic-spin inclination angle, $\alpha$, 
from its initial value $\alpha_0$:
\begin{equation}
    \sin\alpha(t) = \exp(-t/\tau)\,\sin\alpha_0 \,,
    \label{eq:alphat}
\end{equation}
where $t$ is the pulsar's age and $\tau$ is the alignment time-scale. The 
alignment has a simple electromechanical origin: the electromagnetic radiation 
emitted by an oblique rotating dipole results in a torque on the star that 
causes the angular velocity axis to migrate through the neutron star toward 
alignment with the magnetic axis. 

The second effect is progressive narrowing of the emission cone, described by a 
power-law dependence of its half-angle, $\rho$, on the rotation period, $P$:
\begin{equation}
    \rho(t) \propto P^{-\gamma}(t) \,,
\label{eq:rho:t}
\end{equation}
with $\gamma$ a positive constant having a value between $1/3$ and $2/3$ 
\citep{go70,rs75,lm88,ran93}.

For a pulsar with constant rotational inertia and fixed magnetic moment vector, 
the rotation period evolves according to $\dot{P}P^{n-2}$ = constant. It was 
pointed out long ago \citep{pb81} that this simple rule is incompatible with 
the observed distribution of pulsars in the $P$--$\dot{P}$ plane, conflicting 
with stationary `flow' through that plane to pulsar `death' at advanced ages.  
However, alignment produces an effective reduction in the magnetic moment, 
reducing the torque in proportion to $\sin{^2}\alpha$; so, in the Jones model, 
the period evolves according to \citep{pb81}
\begin{equation}
    \dot{P}P^{n-2} \propto \exp(-2t/\tau) \,,
\label{eq:PdotPn}
\end{equation}
with $n$ (the braking index) constant.  

\Eqn{eq:PdotPn} integrates to give:
\begin{equation}
    P^{n-1}(t) = P{_\infty}^{n-1}+ (P{_0}^{n-1}- P{_\infty}^{n-1})\exp(-2t/\tau) \,,
\label{eq:P:t}
\end{equation}
where $P_0$ is the pulsar's initial period and $P_\infty$ is its
limiting period as $t \rightarrow \infty$.  Assuming that pulsars
eventually slow substantially from their initial spin period, $ P_0 \ll
P_\infty$, we can approximate \eqn{eq:P:t} by
\begin{equation}
    P(t) \approx P_{\infty}[\,1 - \exp(-2t/\tau)\,]^{1/(n-1)} \,.
    \label{eq:P:t:approx}
\end{equation}
Using \eqNo{eq:P:t:approx} for $P(t)$ in \eqNo{eq:rho:t} for
$\rho(t)$ yields a simple approximation for the evolution of $\rho$,
expressed directly in terms of $t$:
\begin{equation}
    \rho(t) \approx \rho_{\infty} [\,1 - \exp(-2t/\tau)\,]^{-\gamma /(n-1)} \,,
    \label{eq:rho:t:approx}
\end{equation}
where $\rho_\infty$ is the limiting emission cone half-angle as $t \rightarrow 
\infty$.

We note that, because $P_0$ has effectively been taken as zero in
going from \eqNo{eq:P:t} to \eqNo{eq:P:t:approx} for $P(t)$,
\eqn{eq:rho:t:approx} is not applicable to the very youngest pulsars:
for $t \ll \tau/2$ it reduces to
\begin{equation}
    \rho(t)/\rho_\infty \approx (\tau/2t)^{\gamma/(n-1)} \,, 
    \label{eq:rho:t:approx:2}
\end{equation} 
which diverges as $t \rightarrow 0$.  For the typical parameter values
(see \secref{sec:DataFits}
\tableref{table:parameters:three:nGamma:min}) $\gamma = 1/2,$ $n =
2.3,$ $\tau = 1 \times 10^6\,$yr, and $\rho_\infty = 2-5\degr$,
\eqn{eq:rho:t:approx:2} gives $\rho(t) \approx 11\rho_\infty \approx
22-55\degr$ for $t \approx 1000\,$yr.  As only one known pulsar is this
young, the divergence is outside the age range we can analyse, and the
approximation \eqNo{eq:rho:t:approx} for $\rho(t)$ is valid for our
purposes.

\subsection{Evolution in characteristic age} 

A pulsar's spin-down age, \(\SpinDownAge{}\LRdefined{}P/\left[(n-1)\dot{P}\right]\), 
depends on $n$, which is going to be a variable in our fitting procedure below. 
Instead of the spin-down age, from here on we use the the characteristic age 
which is defined to be
\begin{equation}
    t_{c}\LRdefined{P}/\left({2\dot{P}}\right)
    \mbox{~.}
    \label{eq:CharacteristicAge}
\end{equation}
The characteristic age is independent of $n$ and corresponds to the spin-down age
in the case $n=3$.  Its use enables us to show fittings for different $n$ on one plot 
using the same characteristic-age data.  Inserting \eqn{eq:P:t:approx} for $P(t)$, 
we find that $t_c$ is related to the pulsar's actual age, $t$, by:
\begin{eqnarray}
    t_{c}(t)/\tau 
    & = &
    [\,(n-1)/4\,][\,\exp(2t/\tau) - 1\,]
    \label{eq:tc:t}\\
    \mbox{or~~} 
    \exp[2t(t_c)/\tau] 
    & = & 
    1 + (4t_{c})/[(n-1)\tau]
    \mbox{~.}
    \label{eq:t:tc}
\end{eqnarray}
Note that factors of $2/(n-1)$ effectively rescale $t_{c}$ to
$\SpinDownAge{}$ in these equations.

It follows from \eqn{eq:tc:t} that $t_c \approx (n-1)t/2$ for
relatively young pulsars (young compared with the spin-down
time-scale); for old pulsars, $t_c/\tau$ increases as an exponential
function of $t/\tau$.  The characteristic and actual ages coincide for
relatively young pulsars in the dipole rotator case, for which $n =
3$.

The $t_c$--$t$ relation enables \eqns{eq:alphat}, \eqNo{eq:P:t:approx}
and \eqNo{eq:rho:t:approx} for $\alpha$, $P$ and $\rho$ to be
re-expressed in terms of $t_c$ -- which has the advantage of being
directly determined from the measured quantities $P$ and $\dot{P}$ --
instead of in terms of $t$:
\begin{eqnarray}
    \sin\alpha(t_c) 
    \! & = & \!
    \left(\sin\alpha_0\right)
    \left[1+{4t_c}/{(n-1)\tau}\right]^{-1/2} 
    \mbox{~,}
    \label{eq:alpha:tc} \\
    P(t_c) 
    \! & = & \!
    P_{\infty}\left[1+{(n-1)\tau}/{4t_c} \right]^{-1/(n-1)}
    \mbox{~,}
    \label{eq:P:tc} \\
    \mbox{and}\hspace{5mm} 
    \rho(t_c) 
    \! & = & \!
    \rho_{\infty} \left[1+{(n-1)\tau}/{4t_c} \right]^{\gamma/(n-1)} 
    \mbox{~.}
    \label{eq:rho:tc}
\end{eqnarray}

\Eqn{eq:rho:tc} describes an emission-cone half-width that
decreases quite rapidly with time for young pulsars; consequently, the
observed pulsewidths should decrease with age for the youngest
pulsars.  The alignment of the spin and magnetic axes, described by
\eqn{eq:alpha:tc}, should cause an increase in the observed
pulsewidths for older pulsars, as the pulse will occupy a larger
fraction of a complete rotation.  The combination of these two
processes should lead to a minimum in the observed angular pulsewidths
for pulsars of moderate ages (CB83).

Note that taking the derivative of \eqn{eq:P:t} and using \eqn{eq:CharacteristicAge} for $t_c$ gives (cf. \eqn{eq:BrakingIndex:Definition})
\begin{equation}
	\nAPP{app}=n + 4 {t_{c}}/{\tau}
	\mbox{~,}
	\label{eq:nApp:n:tc} 
\end{equation}
showing that the apparent braking index increases with characteristic age.

\subsection{The parameters of the theory}

\Eqns{eq:alpha:tc}, \eqNo{eq:P:tc} and \eqNo{eq:rho:tc}
-- expressing the evolution of the magnetic-spin inclination angle,
the pulsar period and the emission cone half-angle in terms of a
pulsar's characteristic age -- can conveniently be re-written in terms
of a normalized dimensionless characteristic age:
\begin{equation}
    T_c \LRdefined 2t_c/[(n-1)\tau]\;,
    \label{eq:Tc:defn}
\end{equation}
which reduces to $t_c/\tau$ in the dipole rotator $(n=3)$ case. Thus:
\begin{eqnarray}
    \sin\alpha(T_c) 
    \! & = & \! \left(\sin\alpha_0\right) (1 + 2T_c)^{-1/2}  
    \mbox{~,}
    \label{eq:alpha:Tc} \\
    P(T_c) 
    \!& = &\!
    P_{\infty}[1+(2T_c)^{-1}]^{-1/(n-1)}
    \mbox{~,}
    \label{eq:P:Tc} \\
    \mbox{and~~}
    \rho(T_c) 
    \!& = &\! 
    \rho_{\infty}[1+(2T_c)^{-1}]^{\gamma/(n-1)}
    \mbox{~.}
    \label{eq:rho:Tc}
\end{eqnarray}

In the equations used for the pulsewidth modelling below, namely
\eqNo{eq:alpha:tc} and \eqNo{eq:rho:tc}, or \eqNo{eq:alpha:Tc} and
\eqNo{eq:rho:Tc}, for $\alpha(t_c)$ and $\rho(t_c)$, $n$ appears only
in the two combinations $\gamma/(n-1)$ and $(n-1)\tau$.  Hence there
are only four basic independent parameters in the Candy-Blair
alignment models.  Later we will set $\gamma=1/2$, consistent with the
$\gamma$ values found by other authors
\citep[e.g.][]{ran93,mck93,gou94}.
Hence, it is convenient in this study to define the four independent
parameters to be $\alpha_0$, $\rho_\infty$, $n_{\gamma}$ and
$\tau_{\gamma}$, where
\begin{eqnarray}
    n_{\gamma} & \LRdefined &
    1+\left(n-1\right)/ (2\gamma)\mbox{~,~and}
    \label{eq:parameters:gamma:n}\\
    \tau_{\gamma} 
    & \LRdefined & 
    2\gamma\, \tau  \mbox{~,}
    \label{eq:parameters:gamma:tau}
\end{eqnarray}
since these last two parameters are normalized to the $\gamma=1/2$ case.
In terms of these parameters, the two combinations $\gamma/(n-1)$ and
$(n-1)\tau$ become
\begin{eqnarray}
    \gamma/(n-1) & = & 1/\left[2\left(n_{\gamma}-1\right)\right]
    \mbox{~,~and}\\
    (n-1) \,\tau & = & (n_{\gamma}-1)\tau_{\gamma} \mbox{~,}
    \label{eq:parameters:gamma:tau:2}
\end{eqnarray}
and \eqn{eq:nApp:n:tc} gives
\begin{equation}
   \nAPP{app}-1
   =
   2\gamma \left( n_\gamma -1 +\frac{t_c}{\tau_\gamma} \right)
   \mbox{~.}
   \label{eq:nApp:nGamma:tc}
\end{equation}

\section{PULSEWIDTH EVOLUTION DESCRIPTION}
\label{sec:PulsewidthEvolution}

\subsection{Mean pulsewidths}

For an individual pulsar, the observed angular pulsewidth is given as a function of $\alpha, \rho$ and $\zeta$ by
\begin{equation}
    W(\alpha,\rho,\zeta) 
    = 
    2\arccos{\Big(\frac{\cos\rho - \cos\alpha\cos\zeta} {\sin\alpha \sin\zeta}\Big)} \,,
    \label{eq:W:defn}
\end{equation}
where $\zeta$ is the angle between the observer's direction and the pulsar's spin axis  \citep[][p.~218]{mt77}. The strong dependence on the orientation of the observer with respect to the spin axis leads to a wide scatter in the pulsewidth values, so we need to take a mean pulsewidth for pulsars with certain values of $\alpha$ and $\rho$. 

The mean pulsewidth is obtained by integrating \eqNo{eq:W:defn} over
the angular extent of the emission beam:
\begin{equation}
    \left<{W}\right>\!(\alpha,\rho) = \int_{\mid\alpha-\rho\mid}^{\alpha+\rho}
    W(\alpha,\rho,\zeta) P(\zeta)\mathrm{d}\zeta
    \mbox{~,}
    \label{eq:W:bar}
\end{equation}
in which $P(\zeta)\mathrm{d}\zeta$ is the probability that the angle between the 
spin 
axis and the observer's direction is in the range $\zeta$ to $\zeta + 
\mathrm{d}\zeta$. The integration is over all angles $\zeta$ for which emission is 
both directed toward the observer and seen as pulsed.

\begin{figure}
\includegraphics[width=84mm]{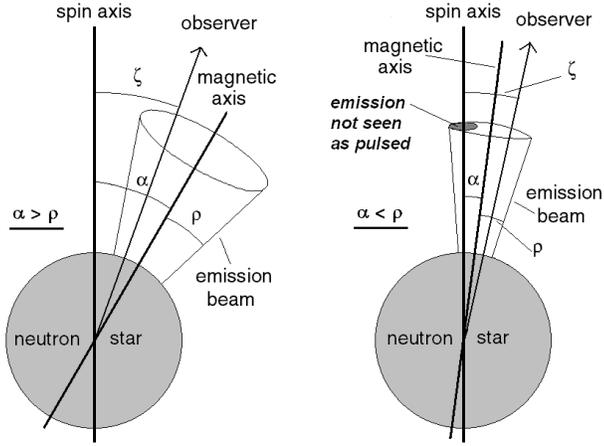}
\caption{Pulsar emission relative to a distant observer, assuming a circular 
cone of emitted radiation, with the magnetic-spin inclination angle $\alpha$ 
either greater (left) or less (right) than the emission cone half-angle $\rho$. 
In the latter case, the emission cone encloses the spin axis, and there may be 
little or no modulation; in particular, to observers in the cone $\zeta < \rho - 
\alpha$, the emission cone is continuously visible. The left-hand diagram 
depicts the usual case of a pulsar with moderate values of $\alpha$ and $\rho$, 
typical of mature-aged pulsars. The right-hand diagram depicts a pulsar with a 
thin beam and nearly aligned spin and magnetic axes, typical of old pulsars in 
the Candy-Blair theory and the Jones model.}
\label{fig:1}
\end{figure}

The limits of integration in \eqNo{eq:W:bar} for
$\left<{W}\right>\!(\alpha,\rho)$ are explained by means of
\figref{fig:1}.  For a pulsar that is not too close to alignment,
such that the emission beam does not contain the rotation axis
(\figref{fig:1}, left), $\alpha > \rho$ and the beam is
intercepted for a part (but not all) of the rotation period if $\zeta$
is in the range $\alpha - \rho$ to $\alpha + \rho$.  For a pulsar that
is close to alignment, with the emission beam containing the rotation
axis (\figref{fig:1}, right), $\alpha < \rho$ and the range $\zeta
< \rho - \alpha$ is excluded, as some part of the beam is then always
directed toward the observer, assuming the emission cone to be filled.

We take $\alpha \leq 90\degr$; the choice of first or second quadrant
for $\alpha$ is a matter of convention, switching it between
hemispheres being equivalent to reversing the sense of the neutron
star's rotation.  The beam, however, can extend into the other
hemisphere, which it will do if $\alpha + \rho > \pi/2$, as expected
for young pulsars with wide beams.
If a pulsar produces observable radio emission from both poles, then 
this will be observed as a main pulse and interpulse, provided that 
the pulsar is nearly orthogonal or the pulsar beams are wide.  However, 
the mean pulsewidths referred to here are for emission from only one 
of these poles, not the combined main pulse -- interpulse width.

On taking pulsar spin axes to be randomly directed with respect to observers, 
and normalizing over the angular extent of the emission cone, the probability 
distribution for $\zeta$ becomes:
\begin{equation}
    P(\zeta)\mathrm{d}\zeta = (\sin{\zeta}\mathrm{d}\zeta / 2)/ 
    f_b(\alpha,\rho)
    \mbox{~,}
    \label{eq:Prob:zeta}
\end{equation}
where $f_b$ is the `beaming fraction', which measures the fraction of
the sky swept out by the beam.  \Eqn{eq:Prob:zeta} is the
ratio of the surface area of a ring on a sphere, corresponding to
observers in the range $\zeta$ to $\zeta + \mathrm{d}\zeta$, to the
surface area of the sphere swept out by the beam.

The surface area of the part of a sphere between co-latitudes $\theta_1$ and 
$\theta_2$, normalized to the surface area of a sphere, is
\begin{equation}
    \sin\left[(\theta_1+\theta_2)/2\right]
    \sin\left[(\theta_2 - \theta_1)/2\right]
    \mbox{~.}
    \label{eq:Sphere:Area}
\end{equation}
Inserting $\theta_1 = \left|\alpha -\rho\right|$ and $\theta_2 = \alpha +\rho$ gives 
the beaming fraction:
\begin{equation}
    f_{b}(\alpha,\rho) = \sin{\alpha} \sin{{\rho}} \,, 
    \label{eq:fb}
\end{equation}
applicable to both situations depicted in \figref{fig:1}. 

We note that \citet[][eq.~(7)]{tm98} use a different beaming fraction
for the case $\alpha < \rho$, namely $1 - \cos(\alpha+\rho)$, which
results from integration over 0 to $\alpha+\rho$.  As they remark,
there may be little or no modulation when $\alpha < \rho$, and we
prefer to exclude the `continuously seen' cone, $\zeta < \rho -
\alpha$ (\figref{fig:1}, right).

\subsection{Evolution modelling}

The evolutionary time dependence of $\alpha$ and $\rho$,
\eqns{eq:alphat} and \eqNo{eq:rho:t:approx}, results in a
mean pulsewidth, $\left<W\right>\!(\alpha,\rho)$, that is a function of
time only, $\left<W\right>\!(t)$.  This function is to be treated as a
density, $\left<W\right>\!(t)\,\mathrm{d}t$ giving the mean pulsewidth
of pulsars aged between $t$ and $t+\mathrm{d}t$.  The age of the
pulsars will here be measured in terms of their log characteristic
age, $\log{t_{c}}$, and will be binned with respect to that variable.
Under the change of variables \(t\rightarrow t(\log t_{c})\),
we have
\begin{eqnarray}
    \left<{W}\right>\!(t) \mathrm{d}t
    & =  &
    \left<{W}\right>\!\left(t(\log{t_{c}})\right) \,
    \frac{\mathrm{d}t}{\mathrm{d}\left(\log{t_{c}}\right)}   
    \mathrm{d}\left(\log{t_{c}}\right)\\
    & \equiv & 
    \left<{W}\right>\!(\log{t_{c}}) f_{t}
    \mathrm{d}\left(\log{t_{c}}\right)
    \mbox{~.}
    \label{eq:ft:definition}
\end{eqnarray}
The factor $f_{t}\left(t_{c}\right)$ is given by
\begin{equation}
    f_{t}
    \LRdefined 
    \frac{\mathrm{d}t}{\mathrm{d}\log{t_c}} = \frac{\tau \ln{10}}{2} 
    \left(1 + \frac{(n-1) \tau}{4 t_{c}} \right)^{-1} \,,
    \label{eq:ft:tc}
\end{equation}
where the $t$--$t_{c}$ relation, \eqn{eq:t:tc}, has been
used.  So $f_{t}\left(t_{c}\right)$ increases monotonically from 
$(\ln{10}) 2 t_{c}/(n-1)$ for young pulsars, flattening to 
$(\ln{10})\tau/2$ for old ones.

To allow for variations between the evolutionary histories of
different pulsars, CB86 and C93 relaxed the assumptions that all
pulsars are born with the same inclination angle and have the same
alignment time-scale, $\alpha_{0}$ and $\tau$ respectively.  Instead,
these are replaced by a probability distribution function,
$P(\alpha_0)$, for initial inclination angles (see
\secref{sec:AlignmentModels}), and a log-normal distribution of
alignment time-scales:
\begin{equation}
    P(\log{\tau}) = \frac{1}{\sqrt{2\pi} \sigma_{\log}} 
    \exp\left[-\frac{1}{2}\left(\frac{\log{\tau} - \mu_{\log}}
    {\sigma_{\log}}\right)^2\right] 
    \mbox{,}
    \label{eq:Prob:log:tau}
\end{equation}
where $\mu_{\log}$ and $\sigma_{\log}$ are the mean and standard
deviation of the logarithm of the time-scale distribution,
$\log{\tau}$.  If the normalized parameters $n_{\gamma}$ and
$\tau_{\gamma}$ that were introduced in
\eqns{eq:parameters:gamma:n}--\eqNo{eq:parameters:gamma:tau:2}
are used, then $\tau_\gamma$ has a log-normal distribution with mean
\begin{equation}
    \mu_{\gamma,\log}  =  
    \mu_{\log} + \log\left({2 \gamma}\right)
    \label{eq:mu:gamma:log}
\end{equation}
and standard deviation $\sigma_{\log}$ unchanged.

The mean observed pulsewidth as a function of characteristic age is 
now given by
\begin{eqnarray}
    \lefteqn{
    \left<{W}\right>\!(\log{t_{c}}) =
    }&&\nonumber\\
    &&
    \frac{1}{A}\int\limits_{0}^{\infty} 
    \int\limits_{0}^{\pi/2} \left<{W}\right>\!(\alpha,\rho)f_{b}f_{t} P(\alpha_0) 
    P(\log{\tau}) \mathrm{d}\alpha_{0} \frac{\mathrm{d}\tau}{\tau} \,,
    \label{eq:Wbar:tc}
\end{eqnarray}
in which the normalization factor, $A$, is 
\begin{equation}
    A(\log{t_{c}}) = \int_{0}^{\infty} \int_{0}^{\pi/2} f_{b}f_{t} P(\alpha_0) P(\log{\tau}) 
    \mathrm{d}\alpha_{0} \frac{\mathrm{d}\tau}{\tau} \,.
    \label{eq:A:normalization}
\end{equation}
The beaming fraction, $f_{b}$, must be included as a weighting factor
in \eqn{eq:Wbar:tc} to take into account the effect of
beaming on the probability of observation, and those weights are
normalized by the factor $A^{-1}$ (the mean pulsewidth is only
computed for those pulsars that are observed).  The aim here is to fit
\eqn{eq:Wbar:tc} for $\left<{W}\right>\!(\log{t_{c}})$ to
the observed data.

In Appendix~\ref{sec:Appendix:RRB} it is shown that
\begin{equation}
    \left<{W}\right>\!(\alpha,\rho) f_{b} 
    =
    \left\{
    \begin{array}{cll}
        \left[1-\cos\rho\right]\pi & \mbox{for} & \alpha \geq 
	\rho\mbox{~,}  \\
        \left[\cos(\rho-\alpha)-\cos\rho\right]\pi & \mbox{for} & 
	\alpha \leq \rho \mbox{~.}
    \end{array}
    \right.
    \label{eq:PWint}
\end{equation}
Inserting this result into \eqn{eq:Wbar:tc} for
\(\left<{W}\right>\!(\log{t_{c}})\) allows a considerable improvement in the
numerical performance of the curve-fitting.

At a given characteristic age, \Eqn{eq:rho:tc} implies that \(\rho(t_{c}) > 90\degr\)
if a pulsar were to have \(\tau>\tau_{\mathrm{cut}}\), where
\begin{eqnarray}
    \tau_{\mathrm{cut}}\left(t_{c}\right) 
    & \LRdefined & 
    \frac{4 t_{c}}{n-1}
    \left[ 
    \left(\frac{90\degr}{\rho_{\infty}}\right)^{(n-1)/\gamma} -1 \right]
    \label{eq:tau:max} \\
    & \approx &
    \frac{4 t_{c}}{n-1} \left( \frac{90\degr}{\rho_{\infty}} \right)^{\frac{n-1}{\gamma}}
    \mbox{,~~for~}
    \rho_{\infty} \ll 90\degr
    \mbox{.}
\end{eqnarray}
This unphysical possibility arises from the approximation made earlier
that \(P_{0}/P_{\infty}\rightarrow{}0\).  In reality a pulsar with
\(\tau > \tau_{\mathrm{cut}}\) must be born with a characteristic age greater
than the $t_{c}$ used in \eqn{eq:tau:max}.  Therefore
integrals over $\tau$ are terminated at an upper limit of $0.99
\tau_{\mathrm{cut}}$, where the factor 0.99 is included to assist numerical
convergence.

\subsection{Three initial alignment models}
\label{sec:AlignmentModels}

We use three different initial alignment models, $P(\alpha_0)$, from
C93 that involve different distributions of the initial inclination
angles.  Model~I assumes that all pulsars start with their spin and
magnetic axes mutually perpendicular: $\alpha_0 = 90\degr$.  This corresponds to the \citeauthor{jon76} model  following the initial brief phase of counter-alignment to orthogonality.   Model~II
uses the probability distribution
\begin{equation}
    P(\alpha_{0}) = (2 - \surd 2)^{-1}\sin(\alpha_{0}/2) \,,
    \label{eq:Prob:alpha0:II}
\end{equation}
which assumes pulsars are born with a random distribution of inclination angles 
from 0 to $90\degr$. Model~III uses 
\begin{equation}
    P(\alpha_{0}) = (4/\pi)\cos^{2}\alpha_{0} \,,
    \label{eq:Prob:alpha0:III}
\end{equation}
which assumes pulsars tend to be born with inclination angles closer to $0$ 
than 
would occur by chance.  At the outset there is no particular reason to think that Model~III 
might be approximately correct; it is included here for contrast with Models~I and II.

\section{FITTING TO DATA}
\label{sec:Data:Fitting}

\subsection{Data selection}
\label{sec:DataSelection}

Both 10~per~cent and 50~per~cent intensity pulsewidth data, $W_{10}$
and $W_{50}$, are used in this analysis of pulsewidth evolution.  The
data were retrieved from the ATNF pulsar
catalogue
\citep{mhth05}, which includes results from
several surveys at around 400 and 1400~MHz.  For this study, all the
binary, millisecond, anomalous X-ray, rotating radio transient (RRAT) and 
globular cluster
pulsars have been removed from the data set, because these pulsars are
considered to follow different evolutionary histories from those of
the more common isolated radio pulsars.  The condition for excluding
millisecond (or low magnetic field) pulsars was chosen to be
$B_{\mathrm{surf}}\le4\times{}10^{9}\,\mbox{G}$.

\begin{figure}
    \includegraphics[width=84mm]{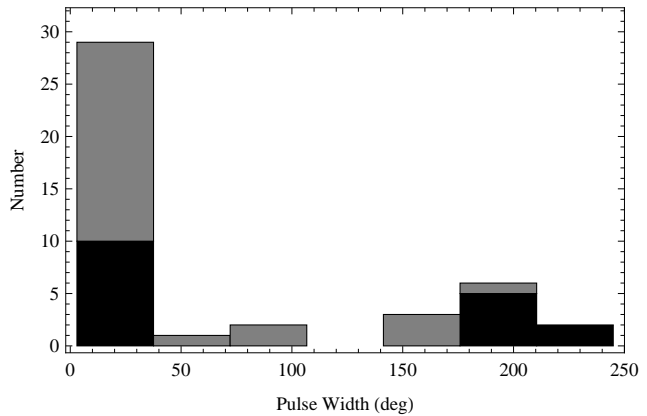}
    \caption{Frequency histograms of the $W_{10}$ (black) and $W_{50}$
    (grey) pulsewidths for the interpulse pulsars listed by \citet{wj08};
    the bars are stacked for clarity.  It is evident that the data
    distributions can be treated as bimodal.  Widths greater than
    $140\degr$ were assumed to have included the interpulses and were
    excluded from the $W_{10}$ and $W_{50}$ pulsewidth data sets used in
    this paper.}
    \label{fig:IP:Histrogram}
\end{figure}

Pulsars with interpulses whose $W_{10}$ or $W_{50}$ values potentially
include both the main pulse and interpulse were also removed.  To do
this the complete census of the interpulse pulsar population recently
published by \citet{wj08} was used.  It identified 27 pulsars with
interpulses, and our \figref{fig:IP:Histrogram} shows histograms of
the existing $W_{10}$ and $W_{50}$ values for them: it is evident that
the distributions can be treated as bimodal with one maximum well below
$90\degr$ and the other at around $180\degr$.  For the purposes of
this study it was assumed that a value of
$\mbox{$W_{10}$~or~$W_{50}$}>140\degr$ includes emission from both
poles of the pulsar (that is both the main pulse and the interpulse),
and so should not be included in the study.  This resulted in 6
pulsars being removed from the $W_{10}$ data set leaving a net total
of 872 pulsars; and 3 interpulse pulsars being removed from the
$W_{50}$ data set, leaving a net total of 1420 pulsars.  These two
data sets are referred to hereafter as `ATNF~Cat~$W_{10}$' and 
`ATNF~Cat~$W_{50}$'.

In a mean pulse profile the spacing between components usually
increases, and with it the whole profile width, as the radio
frequency, $f$, decreases.  For instance, by measuring the profile
widths of 6 pulsars up to 32~GHz, \citet{xkj+96} found that
$W_{50}=b_{0} + b_{1} f^{-p}$ where $p$, $b_{0}$ and
$b_{1}$ are constant for each pulsar, with $0.3<p<0.9$.  The
$W_{10}$ and $W_{50}$ pulsewidths tabulated in the ATNF catalogue are
not all measured at the same observing frequency.  The frequency
dependence of the pulsewidth therefore introduces some statistical
noise into the `ATNF~Cat~$W_{10}$' and `ATNF~Cat~$W_{50}$' data
sets.  As a consistency check, we have also examined the subsets
consisting of only those pulsars whose catalogued pulsewidth values
were measured in the Parkes Multibeam Survey at 1374~MHz
\citep{mlc+01,lfl+06}.  These subsets are
referred to hereafter as `Parkes~MB~$W_{10}$' and `Parkes~MB~$W_{50}$', and
contain 377 and 934 pulsars respectively.

Both the $W_{10}$ and $W_{50}$ characteristic age--pulsewidth data
distributions naturally lend themselves to binning in $\log{t_{c}}$,
and it was also found numerically optimal to fit the mean pulsewidth,
\eqn{eq:Wbar:tc}, as a function of $\log{t_{c}}$ rather
than $t_{c}$, i.e. as $\left<{W}\right>\!(\log{t_{c}})$.  The $W_{10}$
or $W_{50}$ data set is then binned to give $N$ data points,
\(\{\left(\overline{\log t_c}\right)_{i},\overline{W}_{i}\}\),
\(i=1,\ldots,N\).  It is desirable to make the bins as narrow as
possible in order to give maximum resolution in $\log{}t_{c}$.  This
is because if a bin $i$ extends over a large range of $\log{}t_{c}$ in
a region where the underlying function, here $\left<{W}\right>\!(\log
t_{c})$, has a significant slope, then a part of the uncertainty in
the estimated value of the mean pulsewidth, $\overline{W}_{i}$, will
result from the change in the underlying
$\left<{W}\right>\!(\log{t_{c}})$ across the bin.  On the other hand,
it is desirable to include as many points as possible in a bin to
reduce the uncertainty in the $\overline{W}_{i}$ estimate of
$\left<{W}\right>\!\left[\left(\overline{\log t_c}\right)_{i}\right]$,
and also to ensure that the errors in the estimates are approximately
normally distributed so that standard data-analysis procedures can be
applied.  To attempt to balance these competing effects we binned the
data so that each bin contained about 30 pulsars -- the minimum number
of points to ensure that the errors are approximately normally
distributed.

Least-squares fittings are made to both the $W_{10}$ and $W_{50}$ data
sets.  Even though there are more data points in the $W_{50}$ than the
$W_{10}$ data sets, the $W_{10}$ is superior for our pulsewidth
fitting because the $W_{50}$ set suffers more from confusion.  This
is because a $W_{50}$ measurement will include outlying
components in a profile only if they are above 50~per~cent of the maximum
intensity.  A $W_{10}$ measurement is more likely to include the
outlying components and so will give a more consistent estimate of the
total pulsewidth.  A histogram of the ratio of $W_{10}/W_{50}$ is
found to be unimodal, with a peak at about 2.  However, it
extends to ratios above 10, demonstrating that the $W_{10}$ in those
cases is detecting outlying emission missed by the $W_{50}$
measurement.  The trade-off in relying on $W_{10}$ values is that the
$W_{10}$ measurement is more difficult to make than the $W_{50}$, and
so there are fewer pulsars with existing $W_{10}$ measurements.

\begin{figure}
    \includegraphics[width=84mm]{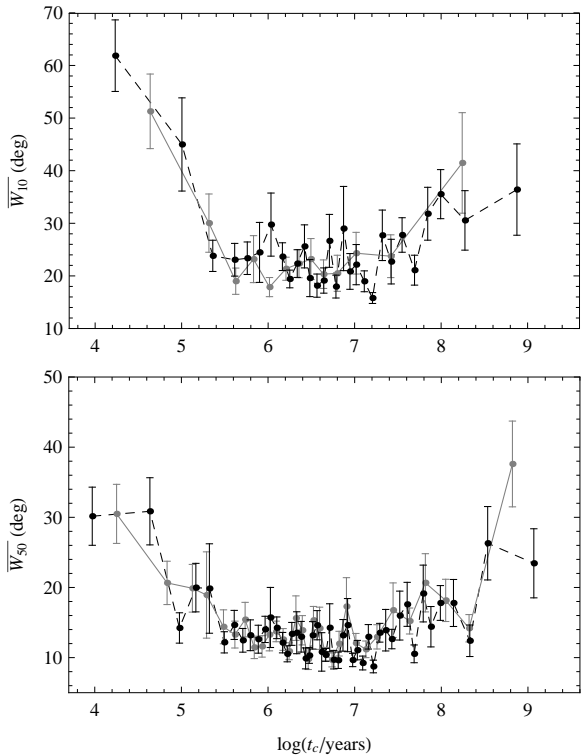}
    \caption{Comparison of the four mean-pulsewidth data sets considered
    in this paper, as summarized in \tableref{table:Datasets}.  The $W_{10}$
    and $W_{50}$ `ATNF~Cat' data sets are shown in black (connected by a 
    dashed
    black line), while the two corresponding `Parkes~MB' data sets are shown
    in grey (connected by a solid grey line).  
    The error bars are 1$\sigma$ standard errors for the binned data.}
    \label{fig:W10:W50:data:comparison}
\end{figure}

The data sets considered in this paper are summarized in
\tableref{table:Datasets}.  For each data set it gives the name of
the set and the number of pulsars it contains, and describes the
binning of the data.  \Figref{fig:W10:W50:data:comparison} shows the
ATNF Cat data sets and compares them to the corresponding Parkes MB
data sets.  We see that the ATNF Cat and Parkes MB data sets have very
similar distributions with respect to characteristic age.

\begin{table}
    \centering
    \caption{Summary of the four data sets considered in this paper.}
    \begin{tabular}{|l|r|l|}
	\hline
	Dataset       & Size & Binning \\
        \hline
        Parkes MB $W_{10}$ &  377 & bins 1--13: 29 pulsars/bin  \\
        Parkes MB $W_{50}$ &  934 & bins 1--14, 19--31: 30 pulsars/bin   \\
	              &      & bins 15--18: 31 pulsars/bin \\
        ATNF Cat $W_{10}$  &  872 & bins 1--14, 17--29: 30 pulsars/bin  \\
	              &      & bins 15--16: 31 pulsars/bin \\
        ANTF Cat $W_{50}$  & 1420 & bins 1--19, 30--47: 30 pulsars/bin  \\
	              &      & bins 20--29: 31 pulsars/bin \\
        \hline
    \end{tabular}
    \label{table:Datasets}
\end{table}

\subsection{Fitting procedure}

Each model of initial alignment has four free parameters:
$\rho_{\infty}$, $n_{\gamma}$, $\mu_{\gamma{},\log}$ and $\sigma_{\log}$.
The aim is to determine these four parameters by least-squares
fitting the model $\left<{W}\right>\!(\log{t_{c}})$-curve to the binned data.
The mean pulsewidths, $\overline{W}_i$, computed for each bin $i$,
have normally-distributed errors, and so the uncertainties for the fitted parameters can
be found by computing the parameter covariance matrix, $s_{k l}$ 
\citep[e.g.][Ch.~15]{ptvf92}.  This matrix can in turn be used to compute a parameter correlation matrix, $r_{k l}$, for the parameter estimates.

Proceeding in this way it was found that least-squares estimates of
the four parameters had large uncertainties, and hence were of
questionable statistical merit.  These large uncertainties were found
to result mostly from a high correlation between the four parameters,
with the off-diagonal values of the parameter correlation matrix all
being around $\pm{}0.98$.  Furthermore, very different sets of
parameter values could give very similar fittings to the data.  

To tackle these problems we fixed $n_{\gamma}$ at specific trial
values, leaving three free parameters: $\rho_\infty$,
$\mu_{\gamma,\log}$, and $\sigma_{\log}$.  The cross-correlation
between the parameters was significantly reduced.  Correspondingly,
the uncertainties in the parameters were also significantly reduced
and, given the assumptions, provided statistically significant
constraints.  The reduced $\chi^{2}$
($\chi^{2}_{\mathrm{red}}\LRdefined\chi^{2}/\mathrm{dof}$) for these 
three-parameter fits were
then examined as a function of the pre-set parameter $n_{\gamma}$; a
minimum in this function would indicate that a particular value of
$n_{\gamma}$ was giving a superior fit.  In that case a least-squares
fit was computed by allowing all four parameters (including
$n_{\gamma}$) to vary, and the solution was checked against the
three-parameter fit for consistency.

We recall that the parameters $n_{\gamma}$ and $\mu_{\gamma,\log}$ are
the values of $n$ and $\mu_{\log}$ assuming that $\gamma=1/2$.  More
generally, for fitted values of $n_{\gamma}$ and $\mu_{\gamma,\log}$,
the relationships between different values $\gamma$, $n$ and
$\mu_{\log}$ are found
via \eqns{eq:parameters:gamma:n} and \eqNo{eq:mu:gamma:log} to be \(n=
2\gamma\left(n_{\gamma} -1\right)+1\), \(\mu_{\log} =
\mu_{\gamma,\log}-\log\left(2 \gamma\right)\), and hence
\begin{equation}
    \frac{n-1}{n_{\gamma} -1} = 10^{\mu_{\gamma,\log}-\mu_{\log}} 
    \mbox{~.}
\end{equation}

\subsection{Fits to the data}
\label{sec:DataFits}

\begin{figure}
    \includegraphics[width=84mm]{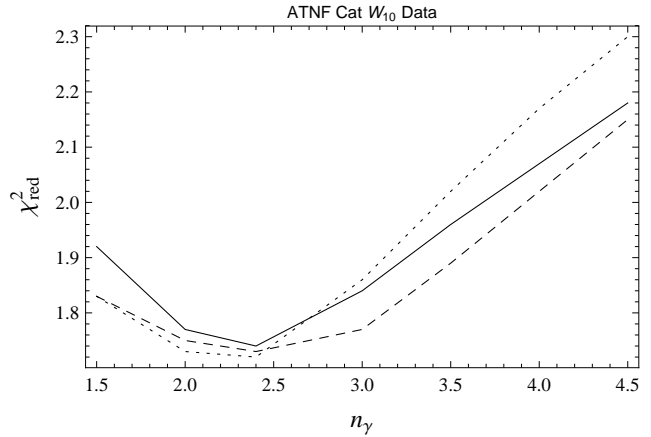}
    \caption{Plot of $\chi^{2}_{\mathrm{red}}$ as a function of
    $n_{\gamma}$, for each of the three models of initial alignment
    fitted to the ATNF~Cat~$W_{10}$ data: Model~I (solid black),
    Model~II (dashed), and Model~III (dotted).  Each model exhibits a
    minimum in $\chi^{2}_{\mathrm{red}}$ at $n_{\gamma}$ around
    2.3.}
    \label{fig:W10:chi2:minima}
\end{figure}

The three-parameter fits showed clear minima in
$\chi^{2}_{\mathrm{red}}\left(n_{\gamma}\right)$ at $n_{\gamma}\approx{2.3}$ 
for all of the datasets, excluding the ATNF~Cat~$W_{50}$ data.
For instance, \figref{fig:W10:chi2:minima} plots
$\chi^{2}_{\mathrm{red}}\left(n_{\gamma}\right)$ for Models~I, II and
III fitted to the ATNF~Cat~$W_{10}$ data.   

\begin{figure*}
    \includegraphics[width=174mm]{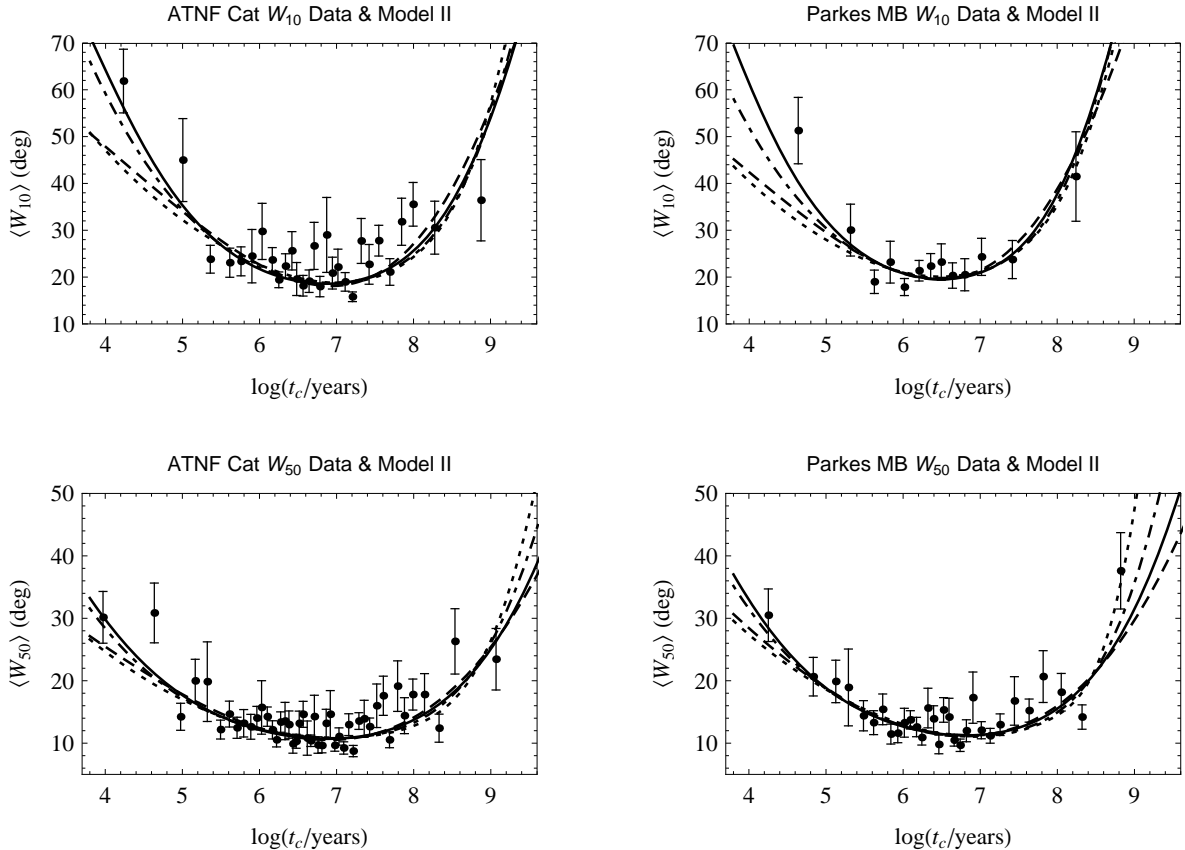}
    \caption{Mean pulsewidth, $\left<{W}\right>$, as a function of log
    characteristic age.  The data sets are given in the title of each
    figure; they are the same as those plotted in \figref{fig:W10:W50:data:comparison} 
    and are summarized in \tableref{table:Datasets}. 
    Error bars are $1\sigma$ standard errors for the binned data. Mean
    angular pulsewidths for Model~II are least-squares fitted to the
    binned data for each of the following values of $n_{\gamma}$:
    $n_{\gamma}=1.5$ (dashed),
    $n_{\gamma}=2.3$ (solid), 
    $n_{\gamma}=3.0$ (dot-dashed),
    and $n_{\gamma}=4$ (dotted).
    The vertical
    scales are different for the $W_{10}$ and $W_{50}$ data sets.  Note that for
    the Parkes MB $W_{50}$ data set, the Model~II curves did not
    converge for $n_{\gamma}=1.5$ or $n_{\gamma}\ge{}4$.  In that
    case, the dashed curve and the dotted curve represent
    $n_{\gamma}=1.7$ and 3.95 respectively.  In each plot, except the
    ATNF Cat $W_{50}$ plot, the solid $n_{\gamma}=2.3$ curve is
    approximately the best-fitting curve to the data across the
    different values of $n_{\gamma}$, having the lowest value of
    $\chi^{2}_{\mathrm{red}}$.  None of the curves in the ATNF Cat
    $W_{50}$ plot was clearly distinguished by a minimum in
    $\chi^{2}_{\mathrm{red}}$.}
\label{fig:Model:II}
\end{figure*}

\Figref{fig:Model:II} shows three-parameter fitted curves of
$\left<{W}\right>\!\left(\log{t_c}\right)$ for Model~II with
$n_{\gamma}$ set at values between 1.5 and 4; equivalent plots for
Models~I and III are similar.  In each case setting
$n_{\gamma}$ to around 2.3 and least-squares fitting the other three
parameters gives the best fit to the data, in particular at low
characteristic age.  Performing the same procedure with $n_{\gamma}$
higher or lower than this value results in a curve that goes to lower
pulsewidth than the measured value at low $\log{t_{c}}$.
The figure shows for the ATNF~Cat~$W_{50}$ data set that even though
$\chi^{2}_{\mathrm{red}}$ does not have a clear minimum at
\(n_{\gamma}\approx{2.3}\), the \(n_{\gamma}\approx{2.3}\) fit appears
to best explain the low-$t_{c}$ mean $W_{50}$ pulsewidths.

\Tableref{table:n} gives the measured values of \nAPP{app} for six pulsars \citep[\othertableref{2},~p.~1291,][and references therein]{lkgk06}.  The uncertainties in the last digit of  \nAPP{app} are shown, being 1$\sigma$ confidence intervals.  The characteristic ages are also listed, indicating that these pulsars are all very young.  Except for the Vela pulsar, the determinations of $\nAPP{app}$ were all made by standard timing analyses that obtained phase coherent solutions \citep{lkg+07}.  This approach could not be used for the Vela pulsar because of its large glitches, so $\dot{\nu}$ was determined from data 150 days after each glitch and then extrapolated back to the glitch epoch; $\ddot{\nu}$ could then be determined from the change in $\dot{\nu}$ over time  \citep{lpgc96}.   

It is noteworthy that the mean value of $\nAPP{app}$ in \tableref{table:n} is $2.42\pm{}0.05$, compared to the values of $n_{\gamma}\approx2.3$ that give the best fits of the Candy-Blair models to the pulsewidth data.  On the grounds of \eqn{eq:nApp:nGamma:tc} we expect $\nAPP{app}\approx{}n_{\gamma}$ for $\gamma\approx1/2$ and $t_c\ll\tau_{\gamma}$.  \Figref{fig:Model:II} shows that the constraint of  $n_{\gamma}\approx{}2.3$ is based primarily on young pulsars, but a much larger sample than the six given in \tableref{table:n}.

\begin{table}
    \caption{Measured values of the apparent braking index,
    $\nAPP{app}$. }
    \begin{tabular}{|l|r|l|l|}
	\hline
	Pulsar 		      & $t_c\,$(yr)       & $\nAPP{app}$ & Reference \\ 
	\hline
	J1846-0258	      & 723 	              & $2.65(1)$        & \citet{lkgk06} \\
	B0531+21, Crab    & 1240 	     & $2.51(1)$        & \citet{lps93}   \\
	B1509-58 	      & 1550 	     & $2.839(3)$      & \citet{lkgm05} \\
	J1119-6127	      & 1610 	     & $2.91(5)$        & \citet{ckl+00}  \\
	B0540-69, LMC     & 1670              & $2.140(9)$      & \citet{lkg05}  \\
	B0833-45, Vela     & 11300  	     & $1.4(2)$           & \citet{lpgc96} \\ 
	\hline
    \end{tabular}
    \label{table:n}
\end{table}

\begin{figure*}
\includegraphics[width=174mm]{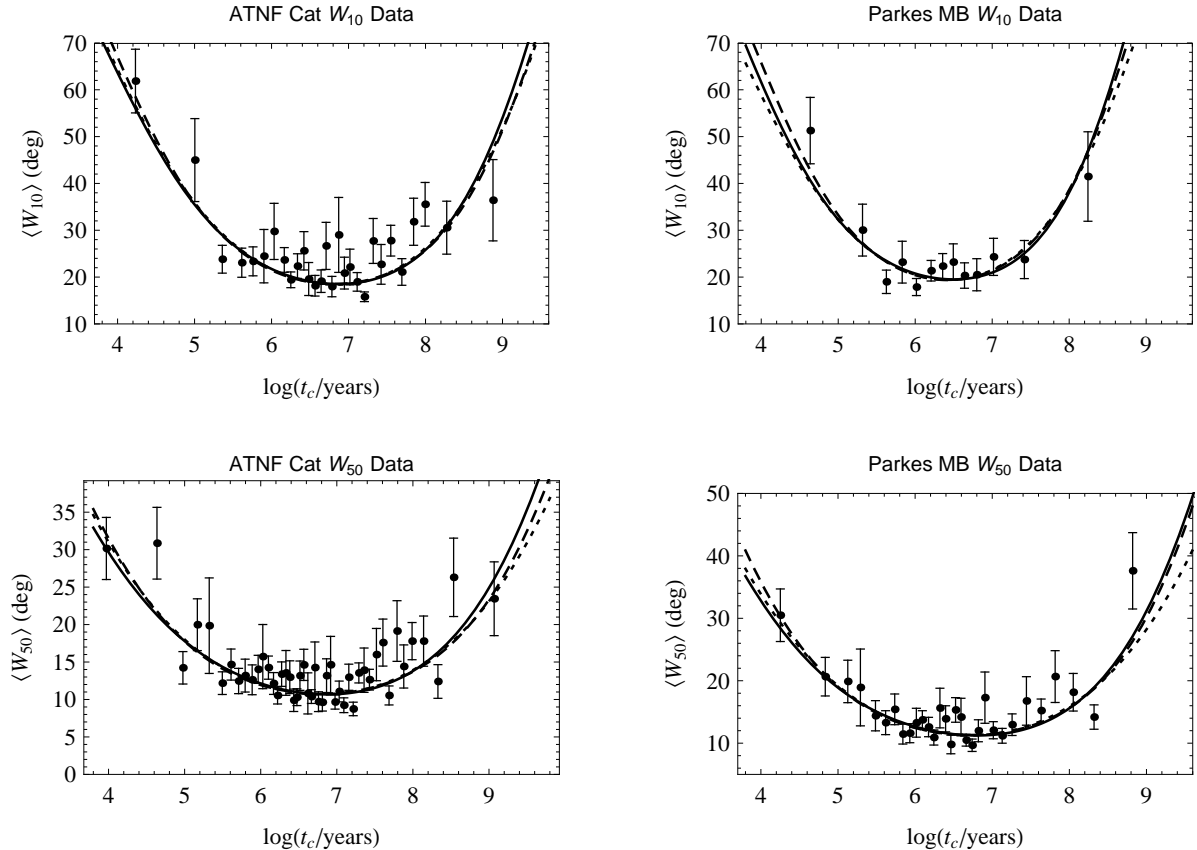}
\caption{Mean pulsewidth, $\left<W\right>$, as a function of log
characteristic age using the fitting parameters given in
\tableref{table:parameters:three:nGamma:min}.  Mean angular
pulsewidths for Models~I (dashed line), II (solid) and III (dotted)
are least-squares fitted to the binned data.}
\label{fig:Models:123}
\end{figure*}

\Tableref{table:parameters:three:nGamma:min} gives the parameter values obtained from the three-parameter fits to the four data sets.  The error ranges are $1\sigma$ confidence intervals, and the degrees of freedom (dof) for each data set are given.  For each fitting the parameter $n_{\gamma}$ is fixed at the value at which $\chi^{2}_{\mathrm{red}}$ is minimum, except in the case of the ATNF Cat $W_{50}$ data where no clear minima were found, and so $n_{\gamma}$ was fixed at 2.3 for comparison with the other fits.  For each data set, all three models predict similar $\mu_{\gamma,\log}$ and $\sigma_{\log}$ values.  The only significant parameter difference between the three alignment models is the predicted limiting emission cone half-width, $\rho_{\infty}$.  \Figref{fig:Models:123} shows curves of $\left<{W}\right>\!\left(\log{t_c}\right)$ for Models~I, II and III using the parameter values listed in\tableref{table:parameters:three:nGamma:min}.   It is evident that the pulsewidth data alone can not be used to distinguish between the three models of initial alignment.

\begin{table}
\caption{Three-parameter least-squares fits of Models~I, II and III.} \begin{tabular}{|l||c|l|l|l|l|l|l|} \\
\hline 
    & $n_{\gamma}$    & $\rho_\infty$ (deg) & $\mu_{\gamma,\log}$ & $
\sigma_{\log}$ & 
${\chi^{2}_{\mathrm{red}}}$\\ 
\hline
\multicolumn{6}{c}{ATNF Cat $W_{10}$ Fits $(\mathrm{dof}=26)$} \\
I   & 2.47            & $5.6\pm{0.2}$       &  $6.5\pm{0.2}$       & $0.71\pm{0.08}$ & 
1.73 \\
II  & 2.30            & $4.11\pm{0.12}$     &  $6.1\pm{0.2}$       & $0.83\pm{0.08}$ & 
1.74 \\
III & 2.25            & $1.99\pm{0.08}$     &  $6.2\pm{0.3}$       & $0.89\pm{0.08}$ & 
1.72 \\
\hline
\multicolumn{6}{c}{ATNF Cat $W_{50}$ Fits $(\mathrm{dof}=44)$}   \\
\multicolumn{6}{c}{$n_{\gamma}\equiv2.3$ (no clear minimum in ${\chi^{2}_{\mathrm{red}}}$)} \\
I   & 2.3 & $2.78\pm0.08$     & $5.4\pm0.3$   & $1.02\pm0.08$   & 1.47  \\
II  & 2.3 & $2.10\pm{0.07}$   & $4.8\pm0.4$   & $1.20\pm0.08$   & 1.46  \\
III & 2.3 & $0.87\pm0.04$     & $5.1\pm0.4$   & $1.26\pm0.09$   & 1.46  \\
\hline
\multicolumn{6}{c}{Parkes MB $W_{10}$ Fits $(\mathrm{dof}=10)$} \\
I   & 2.19            & $5.5\pm{0.3}$       & $6.1\pm{0.4}$       & $0.66\pm{0.13}$ & 
1.05 \\
II  & 2.30            & $4.5\pm{0.2}$       & $6.1\pm{0.4}$       & $0.71\pm{0.14}$ & 
1.18 \\
III & 2.05            & $1.84\pm{0.15}$     & $5.6\pm{0.5}$       & $0.90\pm{0.14}$ & 
1.16 \\
\hline 
\multicolumn{6}{c}{Parkes MB $W_{50}$ Fits $(\mathrm{dof}=28)$} \\
I   & 2.48            & $3.33\pm{0.09}$     & $6.1\pm{0.3}$       & $0.83\pm{0.09}$ & 
1.22 \\
II  & 2.34            & $2.36\pm{0.08}$     & $5.3\pm{0.4}$       & $1.08\pm{0.09}$ & 
1.25 \\
III & 2.20            & $0.87\pm{0.05}$     & $5.1\pm{0.4}$       & $1.21\pm{0.09}$ & 
1.23 \\
\hline
\end{tabular}
\label{table:parameters:three:nGamma:min}
\end{table}

The figures show that following a rapid decline in pulsewidth for
young pulsars, there is a minimum in the pulsewidth for pulsars of
moderate age followed by an increase for older pulsars: this is a
clear signature of spin-magnetic alignment (CB83).  The minimum occurs
at $t_{c}\approx10^{6.9}\,$yr in all three models for all data
sets, except the Parkes~MB~$W_{10}$ data set where it occurs slightly
earlier.  The mean alignment time-scales obtained from the fitting
process are around $10^{6}\,$yr which is less than previous estimates
of around $10^{7}\,$yr.  The log deviations of the time-scales range
from 0.71 to 1.21, and are all larger than the value of 0.25 found by
both CB86 and C93, which suggests that the more sensitive recent
surveys have detected pulsars with a broader range of parameters.

As discussed near the end of \secref{sec:DataSelection}, $W_{10}$ is
superior to $W_{50}$ for pulsewidth fitting.  Hence in the following
we restrict our attention to the $W_{10}$ fits, in particular the
ATNF~Cat~$W_{10}$ fit that was generated from a larger data set.

\section{ANALYSIS AND IMPLICATIONS}
\label{sec:analysis}

\subsection{Magnetic field decay model}
\label{sec:MagneticFieldDecay}

As an alternative to magnetic alignment, many authors have proposed pulsar evolution through magnetic field decay. This will have a different effect on pulsewidth evolution. Suppose that the magnetic field decays exponentially, according to
\begin{equation}
    B = B_{0}\exp(-t/\tau_d) \,,
    \label{eq:B:decay}
\end{equation}
where $\tau_d$ is the field decay time-scale, and that $\alpha=\alpha_{0}$ remains constant (no alignment).  \Eqn{eq:PdotPn} for $P(t)$ still applies, so we still have \eqns{eq:rho:t:approx} and \eqNo{eq:rho:tc} for $\rho(t)$ and $\rho(t_c)$.

\begin{figure}
\includegraphics[width=84mm]{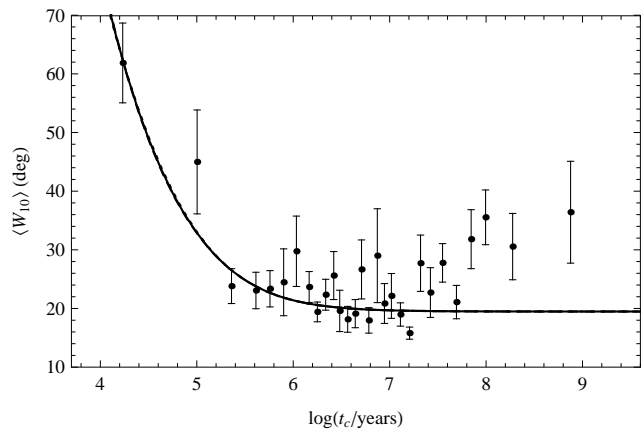}
\caption{The mean angular pulsewidths obtained from the magnetic field
decay model, as fitted to binned ATNF Cat $W_{10}$ data, are plotted
against characteristic age.  It has been assumed here that each pulsar
has a constant inclination angle, $\alpha_{0}$, with the distribution
of those angles determined by either Model~I (dashed), II (solid) or
III (dotted) of inclination angle as outlined in
\secref{sec:AlignmentModels}; the three curves are indistinguishable.}
\label{fig:Magnetic:Decay}
\end{figure}

\Figref{fig:Magnetic:Decay} shows curves of $\left<{W}\right>\!(\log{t_{c}})$
for a constant $\alpha=\alpha_{0}$ (using Models~I, II
and III for the $\alpha_{0}$ distribution) to illustrate the effect of
magnetic field decay on pulsewidth evolution.  The figure shows the
same ATNF Cat $W_{10}$ binned data as before and the curves have been
fitted assuming that $\log{\tau_{d}}$ has a normal distribution
with mean $\mu_{d}$ and standard deviation $\sigma_{d}$.  It was found
necessary to fix $\sigma_{d}$ in order to get a convergent fit, and
the small value of $\sigma_{d}=0.1$ was required in order to increase
the size of $\mu_{d}$ to realistic values.

Qualitatively, the effect of field decay alone on pulsewidth evolution
is a progressive decrease in pulsewidth over time, as seen in
\figref{fig:Magnetic:Decay}.  The data do not support this
evolution path: in contrast to the alignment models, the field decay
model does not account for the increase in pulsewidth values for older
pulsars.

From this work we cannot set an upper limit on the magnetic field
decay rate.  Clearly, weak magnetic-field decay could still occur
concurrently with the alignment mechanism that causes the observed
increase in pulsewidth for old pulsars.  We can say that magnetic
field decay is not the dominant evolutionary factor: both the $W_{50}$ and
$W_{10}$ fits are consistent with magnetic field decay not being a
significant factor in pulsar evolution.  To model the combined effects
of alignment and field decay is beyond the scope of this paper because
of the relatively strong parameter cross-correlation that already
exists in the alignment model alone.

\subsection{Inclination angle evolution}
\label{sec:Alpha:Evolution}

To investigate the question of the alignment of the magnetic and spin
axes, \citet{tm98} used the data sets of \citet{ran93b} and
\citet{gou94} that give estimates of $\alpha$ for a few hundred
pulsars.  These two datasets are derived using slightly different
methods and assumptions, as is well summarized by \citet{tm98}.
\citeauthor{tm98} binned each data set in $\alpha$ to examine the
\(\left<\log{t_{c}}\right>\!(\alpha)\) dependence.  To these they fitted
straight lines (their \otherfigref{6}, p.~632) in order to estimate the
alignment time-scale, which they found to be around
$10^{7}\,$yr.

\begin{figure}
    \includegraphics[width=84mm]{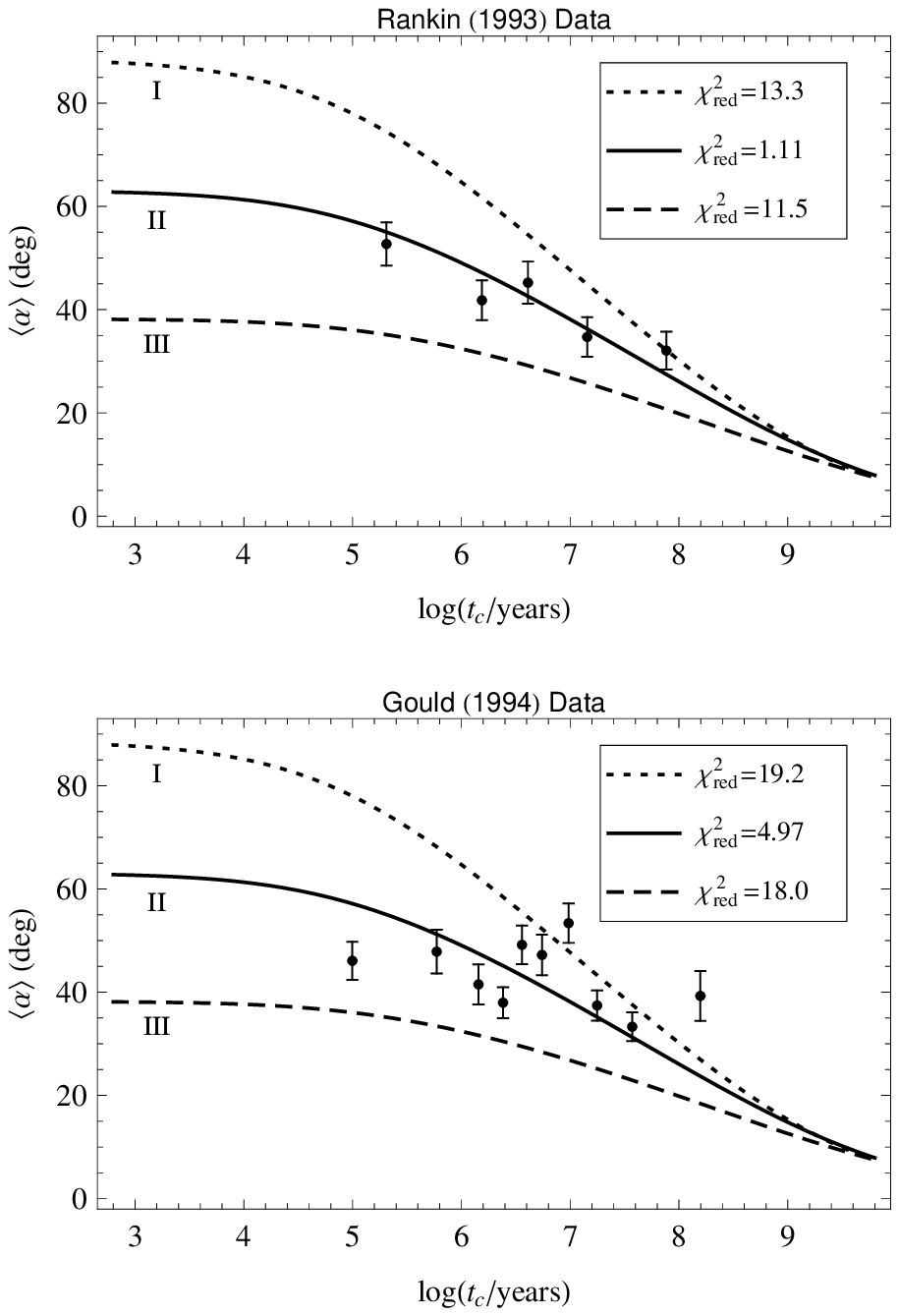}
    \caption{Mean angle of inclination, $\left<{\alpha}\right>$, as a
    function of log characteristic age.  The data are determined from
    models of beam morphology and polarization data and are taken from
    the two references considered by \citet{tm98}.  Binned data are
    shown with $1\sigma$ error bars, and exclude low-field
    (millisecond) pulsars having
    \(B_{\mathrm{surf}}\le4\times{}10^{9}\,\mbox{G}\).  The top figure
    shows the \citet[][\othertablerefs{2 and 4--8}]{ran93b}
    inclination angles for 148 pulsars; the central 3 bins each
    contain 30 points, while the first and last bins contain 29.  The
    bottom figure shows the \citet[][\othertablerefs{1--6}]{gou94}
    inclination angles for 301 pulsars; of the 10 bins, all contain 30
    points, except the sixth bin which has 31.  The plotted curves are
    $\left<\alpha\right>\!\left(\log{t_{c}}\right)$,
    \eqn{eq:alpha:expected}, for each of the three models of initial
    alignment (I, II or III), using the parameter values determined
    from the best fits to the ATNF~Cat~$W_{10}$ data.  The goodness-of-fit
    of the model curves to the binned data is measured using the
    reduced chi-squared statistic, $\chi^{2}_{\mathrm{red}}$, and the
    values for each curve are shown on the plots.  Models~I and III
    are seen to give a poor fit to both the
    \citeauthor{ran93b} and \citeauthor{gou94} mean inclination-angle
    data, while Model~II gives a particularly good fit to the
    \citeauthor{ran93b} data.}
    \label{fig:Alpha:Data}
\end{figure}

Conversely, here the mean inclination angle associated with a 
log characteristic age is found by averaging
the $\arcsin$ of \eqn{eq:alpha:tc}, for 
$\sin\alpha$, over the $P(\alpha_0)$
and $P(\log{\tau})$ distributions:
\begin{eqnarray}
    \lefteqn{
    \left<\alpha\right>\!\left(\log{t_{c}}\right)=
    }&&\nonumber\\
    &&
    \frac{1}{A}\int\limits_{0}^{\infty} 
    \int\limits_{0}^{\pi/2} 
    \alpha(\log{t_{c}})
    f_{b}f_{t} P(\alpha_0) 
    P(\log{\tau}) \mathrm{d}\alpha_{0} \frac{\mathrm{d}\tau}{\tau}
    \mbox{.}
    \label{eq:alpha:expected}
\end{eqnarray}
Using the fitted-parameter values found in
\secref{sec:Data:Fitting} from the various pulsewidth data sets,
the resulting $\left<\alpha\right>\!\left(\log{t_{c}}\right)$ curves for
Models~I, II and III are plotted in \figref{fig:Alpha:Data}.

Also plotted are the data from \citet{ran93b} and \citet{gou94}, in
each case binned with about 30 points per bin.  The error bars are
$1\sigma$ standard errors computed from the binning.  The actual
errors in each of the $\alpha$ values are difficult to evaluate, since
they depend on the validity of the underlying assumptions, as
discussed by \citet{mhq98}.  As with the pulsewidth data, low-field
(millisecond) pulsars having
\(B_{\mathrm{surf}}\le4\times{}10^{9}\,\mbox{G}\) have been removed
from the data.

The model curves are compared with the binned data using the reduced
chi-squared statistic, $\chi^{2}_{\mathrm{red}}$, which is shown for
each model curve in \figref{fig:Alpha:Data}.  It was found that
the inclination angle data are inconsistent with Models~I and III, but
compatible with Model~II. Indeed Model~II predicts the
\citet{ran93b} data particularly well, with a
${\chi^{2}_{\mathrm{red}}}\sim{1}$.

\subsection{Beaming fraction evolution}

From \eqns{eq:alpha:tc} and \eqNo{eq:rho:tc} for 
$\sin\alpha(t_c)$ and $\rho(t_{c})$, \eqn{eq:fb} 
for $f_{b}$ becomes
\begin{eqnarray}
    \lefteqn{
    f_{b}\left(t_{c}\right) 
    = 
    \sin\alpha_0
    \left[1+{4t_c}/{(n-1)\tau}\right]^{-1/2}  
    }&&\nonumber\\
    && 
    \times \sin\left\{\rho_{\infty} \left[1+{(n-1)\tau}/{4t_c} 
    \right]^{\gamma/(n-1)} \right\}
    \mbox{~.} 
    \label{eq:fb:tc}
\end{eqnarray}

\begin{figure}
    \includegraphics[width=84mm]{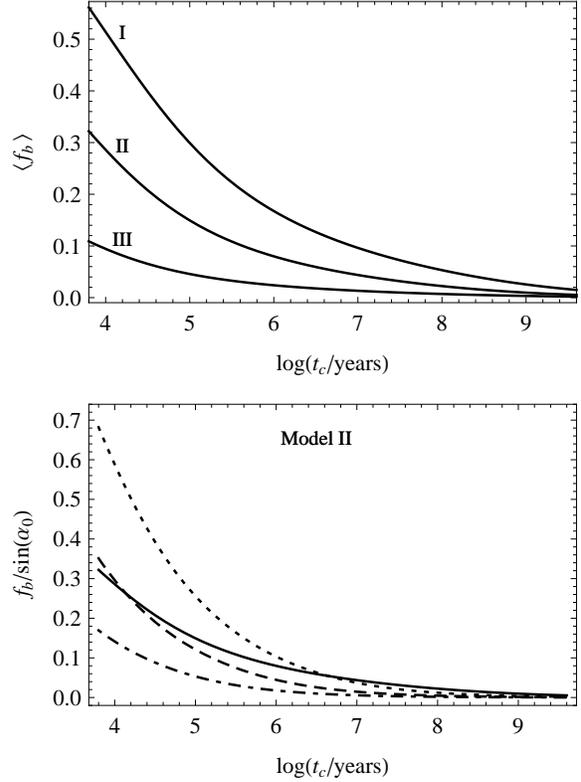}
    \caption{(top) The mean beaming fraction, $\left<f_{b}\right>$, is
    plotted against log characteristic age for Models~I, II and III, as
    labeled.  (bottom) The scaled beaming fraction,
    $f_{b}/\sin\alpha_{0}$, is plotted against log characteristic age
    for Model~II. The dashed curve has
    $\tau_{\gamma}=\mu_{\gamma,\log}$, the dot-dashed curve has
    $\tau_{\gamma}=\mu_{\gamma,\log}-\sigma_{\log}$, and the dotted
    curve has $\tau_{\gamma}=\mu_{\gamma,\log}+\sigma_{\log}$.  For
    comparison, the corresponding Model~II mean beaming fraction from
    the top plot is shown as a solid line.  Both plots use the the
    ATNF~Cat~$W_{10}$ best-fitting parameters,
    \tableref{table:parameters:three:nGamma:min}.}
    \label{fig:fb:mean:scaled}
\end{figure}

The mean beaming fraction as a function of characteristic age,
$\left<{f_b}\right>(t_c)$, is a key quantity in pulsar population
studies.  It is found by integrating $f_{b}\left(t_{c}\right)$ over
the distribution functions $P(\alpha_0)$ and $P(\log\tau)$.  Using the
parameter values from the best fit to the ATNF~Cat~$W_{10}$ data set,
$\left<{f_b}\right>(t_c)$ is plotted in \figref{fig:fb:mean:scaled}
for the three initial alignment models.

\Figref{fig:fb:mean:scaled} shows strong differences in the
$\left<{f_b}\right>(t_c)$ among the three models for young pulsars
because of the differing initial conditions, with gradual reduction to
similar values for older pulsars.  The figure implies that the
smallness of the observed sample size for old pulsars is significantly
influenced by a low probability of observation.  Because of
completeness issues, and luminosity evolution uncertainty, we cannot
evaluate the relative significance of beaming fraction evolution and
pulsar `death' (through reducing radio emission) in determining the
number of observable old pulsars.

For individual pulsars, the beaming fraction evolution will depend on
the pulsar's $\alpha_{0}$ and $\tau_{\gamma}$ values, but the scaled
beaming fraction ${f_b}(t_c)/\sin\alpha_{0}$ is independent of
$\alpha_{0}$.  It is plotted in \figref{fig:fb:mean:scaled} for the
Model~II fit to the ATNF~Cat~$W_{10}$ data, using various values of
$\tau_{\gamma}$.  It is note-worthy that the scaled beaming fraction is
considerably larger for pulsars with a longer alignment time-scale, and
so these pulsars are more likely to be observed than those with a
small alignment time-scale, even at young ages.

\subsection{Alignment time-scales}

From \tableref{table:parameters:three:nGamma:min}, the best-fitting
mean log alignment time-scale, $\mu_{\gamma,\log}$, for Model~II is
found to be $6.1\pm{0.2}$ for the ATNF Cat $W_{10}$ fit, $6.1\pm{0.4}$
for the Parkes MB $W_{10}$ fit, and ${5.3\pm{0.4}}$ for the Parkes MB
$W_{50}$ fit.  The first and last of these three fits have the largest
numbers of degrees of freedom, and they agree within two standard
deviations of error.  Furthermore, a spread in alignment
time-scales is needed to accurately model the data (CB86), and for the
three fits just mentioned, the standard deviations, $\sigma_{\log}$,
are $0.83\pm{0.08}$, $0.71\pm{0.14}$ and $1.08\pm{0.09}$ respectively,
indicative of quite a large spread in alignment time-scales about the
means.

The \emph{mean} time-scales quoted above are smaller than previous
values obtained.  The analyses of CB83, CB86 and C93 yielded a mean
alignment time-scale of $2\times10^{7}\,$yr.  Other authors
have argued in favour of magnetic alignment using different methods
and obtained similar alignment time-scales of around
$10^{7}\,$yr.  See for examples
\tableref{table:alignment:timescales}: \citet{lm88}, \citet{xw91},
\citet{kw92a} and \citet{tm98} all used polarization data, while
\citet{wj08} used interpulse pulsar statistics.

\begin{table}
\caption{Alignment time-scales from different authors.}
\begin{tabular}{|l|r|} \\ \hline
Authors & Alignment time-scale (yr) \\ \hline
CB83, CB86, C93 & $2 \times 10^7$\\
\citet{lm88}  & $10^7$\\
\citet{xw91}  & $1.5 \times 10^7$\\
\citet{kw92a} & $2 \times 10^7$\\
\citet{tm98}  & $10^7$\\ 
\citet{wj08}  & $7\times10^{7}$\\
\hline
\end{tabular}
\label{table:alignment:timescales}
\end{table}

\begin{figure}
    \includegraphics[width=84mm]{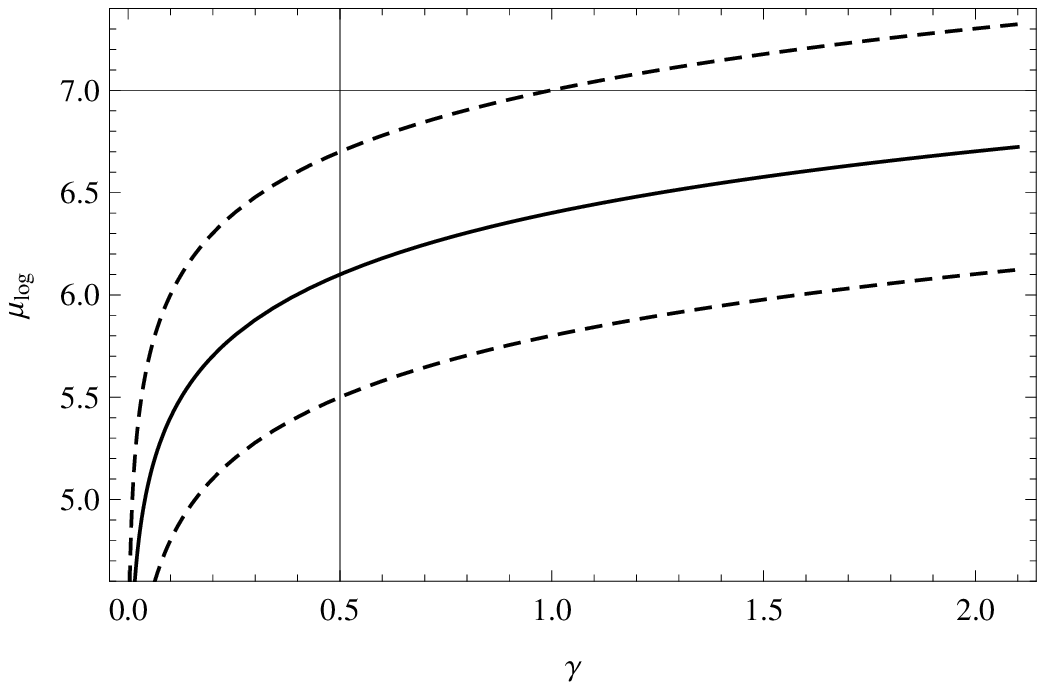}
    \caption{Plots of \eqn{eq:mulog:mugammalog:gamma}: the mean log
    alignment time-scale, $\mu_{\log}$, as a function of the parameter
    $\gamma$, which was introduced in \eqn{eq:rho:t}.  The solid curve
    corresponds to $\mu_{\gamma,\log}=6.1$ from the ATNF Cat $W_{10}$
    Model~II fit (\tableref{table:parameters:three:nGamma:min}), and
    the dashed curves are three standard deviations of error away from
    the black curve, $\mu_{\gamma,\log}=6.1\pm0.6$.  Previous studies
    have found a log alignment time-scale $\ga{}7$.  This is
    inconsistent here with the empirical finding that
    $\gamma\approx{1/2}$.}
    \label{fig:muLog}
\end{figure}

From \eqn{eq:mu:gamma:log}, the fitted value of
$\mu_{\log}$ is determined by the fitted value of $\mu_{\gamma,\log}$
and the value of $\gamma$:
\begin{equation}
    \mu_{\log}=\mu_{\gamma,\log}-\log\left({2 \gamma}\right)
    \mbox{~,}
    \label{eq:mulog:mugammalog:gamma}
\end{equation}
as plotted in \figref{fig:muLog}.  We see that a mean alignment
time-scale $10^{\mu_{\log}}\ga{}10^{7}\,$yr requires $\gamma>1$;
that is inconsistent with previous findings that $\gamma$ is
between $1/2$ and $2/3$.  Conversely, for that range of $\gamma$-values, 
the curves in \figref{fig:muLog} give a mean
alignment time-scale of around $10^{6}\,$yr.

\subsection{Pulsar aging}
\label{sec:PulsarAging}

\begin{figure}
    \includegraphics[width=84mm]{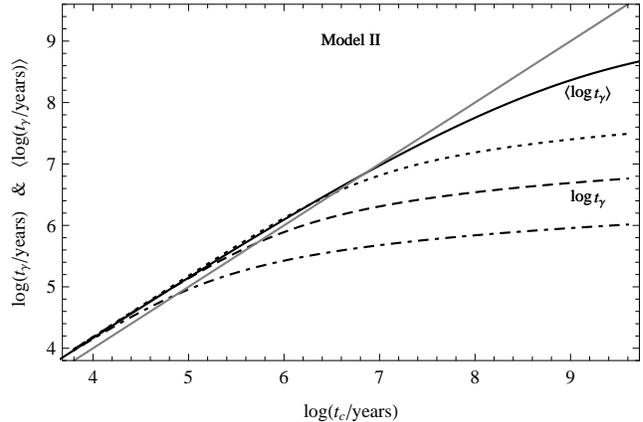}
    \caption{Plots of 
    mean log age, $\left<\log{t_{\gamma}}\right>$, (solid curve) and log age, 
    $\log{t_{\gamma}}$, (broken curves) as functions of log    
    characteristic age for Model~II: 
    (dashed) $\tau_{\gamma}=\mu_{\gamma,\log}$, 
    (dot-dashed) $\tau_{\gamma}=\mu_{\gamma,\log}-\sigma_{\log}$; and
    (dotted) $\tau_{\gamma}=\mu_{\gamma,\log}+\sigma_{\log}$.
    The parameters are from the best fit to the ATNF Cat $W_{10}$
    data set (\tableref{table:parameters:three:nGamma:min}), and the
    straight line is $\log{t_{\gamma}}=\log{t_{c}}$.}
    \label{fig:logt}
\end{figure}

Caution is required in using the characteristic age as a measure of
the true ages of pulsars, as will now be explored.  Defining
$t_{\gamma}\LRdefined 2\gamma t$ and using
\eqns{eq:parameters:gamma:tau} and
\eqNo{eq:parameters:gamma:tau:2} involving $\tau_{\gamma}$, it
follows from \eqn{eq:t:tc} for $t(t_{c})$ that
\begin{equation}
    t_{\gamma}\left(t_{c}\right)
    =
    \frac{\tau_{\gamma}}{2}
    \ln\left[1+\frac{4 
    t_{c}}{\left(n_{\gamma}-1\right)\tau_{\gamma}}\right]
    \mbox{~.}
    \label{eq:tGamma:tc}
\end{equation}
The mean log age, $\left<\log{t_{\gamma}}\right>$,
as a function of the log characteristic age, $\log{t_{c}}$, is found
by integrating over the distribution functions $P(\alpha_0)$ and
$P(\log{\tau})$.  Consider the best fit to the ATNF~Cat~$W_{10}$ data
having parameters as listed in
\tableref{table:parameters:three:nGamma:min}.  A plot of
$\left<\log{t_{\gamma}}\right>\!(\log{t_{c}})$ for Model~II is
shown in \figref{fig:logt}; it shows that  that $\left<\log{t_{\gamma}}\right>$ and $\log{t_{c}}$ agree up
to $\log{t_{c}}\sim{}10 \mu_{\gamma,\log}$, after which time
$\left<\log{t_{\gamma}}\right>$ starts to fall below $\log{t_{c}}$.  The curves for Models~I and III are similar.

However, the mean value $\left<\log{t}\right>\!(\log{t_{c}})$ applies
to the population as a whole.  For individual pulsars, the plots of
$\log{t}(\log{t_{c}})$ depend on the value of $\tau_{\gamma}$ for
that pulsar.  Various tracks for individual pulsars are shown in
\figref{fig:logt}: it is evident that in general the
characteristic age diverges much more rapidly from the true age than
for the populations mean ages, with the divergence being more rapid
the smaller the alignment time-scale.  Recall, however, that pulsars
with shorter alignment time-scales have a smaller beaming fraction than
those with longer time-scales, and so are less likely to be observed.

This highlights the uncertainty that may be present in previous
studies that have used the characteristic age as a measure of true age
when attempting to measure the alignment time-scales.  The
characteristic age can over-estimate the true age, and so will tend to
give larger alignment time-scales if it is the time measure.

\section{CONCLUSIONS}
\label{sec:Conclusions}

From \figref{fig:Models:123}, showing mean pulsewidths versus $\log{t_c}$, we have seen that the pulsewidth data clearly favour pulsar evolution through magnetic alignment, and not magnetic field decay alone nor progressive counter-alignment. There is a clear minimum of mean pulsewidth at about $t_c \approx 10^{6.9}\,$yr. The alignment process causes a significant increase in the mean $W_{10}$  pulsewidth for all characteristic age bins greater than $3 \times10^{7}\,$yr. This is particularly significant in view of the fact that the ATNF catalogue contains numerous pulsars with $t_c > 10^8\,$yr, which were not available to the earlier studies. 

The three models of initial alignment that were specified in \secref{sec:AlignmentModels} cannot be tested by fitting the Candy-Blair models to the pulsewidth data alone.  However, once a Candy-Blair model has been fitted to the pulsewidth data, it can be used to \emph{predict} mean inclination angle evolution, $\left<\alpha\right>\!(\log{t_c})$. Comparing these predictions to the datasets of \citet{ran93b} and \citet{gou94}, it was found that only Model~II (a random initial inclination angle) gives a good fit to the data, as shown in \figref{fig:Alpha:Data}.

The alignment time-scale found from fitting Model~II to the ATNF~Cat~$W_{10}$ data (see \tablerefs{table:Datasets} and \ref{table:parameters:three:nGamma:min}) is $10^{6.1}\,$yr, which is at least a factor of ten shorter than the timescales determined by a number of other authors (see \tableref{table:alignment:timescales}). The fit requires that the log alignment time-scale varies with a standard deviation of about $0.83$ across the observed population.  The fit gives a limiting cone opening half-angle of $4\fdg11$, and a braking index parameter $n_{\gamma}$ of 2.30, consistent with the mean, $2.42\pm{}0.05$, of the measured apparent braking indices (\eqns{eq:BrakingIndex:Definition} and \eqNo{eq:nApp:nGamma:tc}, and \tableref{table:n}).  Consistent with empirical observations, the fit assumes the parameter $\gamma=1/2$ (see \eqn{eq:rho:t}).  For other values of $\gamma$ the fitted parameters $n$ and $\mu_{\log}$ are determined from \eqns{eq:parameters:gamma:n} and \eqNo{eq:mu:gamma:log} respectively.  A plot of the mean beaming fraction for this fit is given in \figref{fig:fb:mean:scaled}.

The initial predictions of pulsar alignment due to electromagnetic
braking \citep{dg70,mg70} predicted time-scales of around
$10^{3}$--$10^{4}\,$yr, which was rather small.  The
time-scales will be increased if pulsars are born with their magnetic
and rotation axes nearly orthogonal.  \citeauthor{jon76}
\citeyearpar{jon76,jon76b,jon76a} found that the inclusion of a
decreasing dissipative torque initially brought all pulsars close to
counter-alignment, from which time the normal electromagnetic decay of
$\alpha$ would begin.  \citeauthor{jon76} predicted that the alignment
phase started at around $10^{3}$--$10^{4}\,$yr, with a
time-scale of around $10^{6}\,$yr, which was believed to be
more in accord with the observed population.

We have found alignment time-scales consistent with the predictions of
\citeauthor{jon76}, though with a significant spread in values, down
to the time-scales consistent with the absence of the
counter-alignment phase.  Contrary to the model of Jones, though, we
do not find evidence for all pulsars starting their alignment phase
with $\alpha_{0}$ close to $90\degr$ (i.e. Model~I).  This interesting
counter-point to the model of Jones deserves further investigation.

\section*{ACKNOWLEDGMENTS}
We thank the Australia Telescope National Facility for making their
pulsar database available.  We thank R.~N.~Manchester for many helpful
discussions and comments on the manuscript, and the referee of
an earlier version of this paper for a helpful review.

\appendix

\section{PULSEWIDTH AVERAGING INTEGRAL}
\label{sec:Appendix:RRB}

The integral that averages the pulsewidth over a random distribution
of viewing angles can be evaluated analytically, as follows.  As
above, $\alpha$ is the pulsar magnetic-spin inclination angle, $\rho$
the emission cone half-width, and $\zeta$ is the angle between the
observer's direction and the pulsar's spin axis.

\subsection {Observer dependence of pulsewidth}

For an individual pulsar, the variation of its observed angular
pulsewidth with observer viewing angle is given by the formula
\citep[][p.~218]{mt77}
\begin{equation}
    W(\alpha,\rho,\zeta) 
    = 
    2\arccos\left( \frac{\cos\rho - \cos\alpha\,\cos\zeta}{\sin\alpha\,\sin\zeta} \right) 
    \,\mbox{,}
    \label{eq:A1}
\end{equation}
which is \eqn{eq:W:defn} in the main text above. 

In considering the variation of $W$ with viewing angle, it is helpful
to employ the variable $x \LRdefined \cos\zeta$.  \Eqn{eq:A1}
takes the form
\begin{equation}
    W(\alpha,\rho,x) = 2\arccos\!\left[f(x)\right] 
    \,\mbox{,}
    \label{eq:A2}
\end{equation}
with the abbreviations
\begin{equation}
    f(x) \LRdefined \frac{1}{\sin\alpha} 
    \frac{b-ax\;}{(1-x^2)^{1/2}}
    \;\mbox{,~}
    a \LRdefined \cos\alpha
    \;\mbox{,~and~}
    b \LRdefined \cos\rho
    \;\mbox{.}
    \label{eq:A3}
\end{equation}
Note that
\begin{eqnarray}
    (\sin\alpha)\,f'(x) 
    & = & (bx-a)/(1-x^2)^{3/2} 
    \quad\mbox{and}
    \label{eq:A4} \\
    (\sin\alpha)\,f''(x) 
    & = &
    (b - 3ax + 2bx^2)/(1-x^2)^{5/2} 
    \;\mbox{.}
    \label{eq:A5}
\end{eqnarray}

The function $f(x)$, which is just the cosine of the pulse angular
half-width, is $< 1$ throughout the cut across the emission beam by
the sight line to any `illuminated' observer.  It reaches $1$ in
magnitude (i.e. $f^2 = 1$) at the pulse visibility limits $\zeta =
|\,\alpha \mp \rho\,|$, corresponding to $x = x_\pm$, where
\begin{eqnarray}
    x_\pm & \LRdefined & \cos(\alpha \mp \rho) = ab \pm \Delta 
    \hspace{4mm}{\mbox{with}}
    \label{eq:A6} \\
    \Delta & \LRdefined & (1-a^2)^{1/2}(1-b^2)^{1/2} 
    \;\mbox{.}
    \label{eq:A7}
\end{eqnarray}

Substituting \eqNo{eq:A6} into \eqNo{eq:A3} gives
\begin{equation}
    f(x_\pm) = \frac{b{}(1-a^2)^{1/2} \mp a{}(1-b^2)^{1/2}}{[1-(ab\pm
\Delta)^2]^{1/2}} 
    \;\mbox{.}
    \label{eq:A8}
\end{equation}
This shows that at $x_-$, the `outer' visibility limit, $f$ is always
positive, i.e. $+1$.  At $x_+$, the `inner' visibility limit, $f$ is
$+1$ for $\alpha > \rho$, $0$ for $\alpha = \rho$, and $-1$ for $\rho
> \alpha$.  Let's visualize these three cases in turn.

(1) For $\alpha > \rho$, the visibility limits $x_\pm$ both correspond
to zero pulsewidth, $W(x) = 0$.  \Eqns{eq:A4} and
\eqNo{eq:A5} show that the minimum value
$(b^2-a^2)^{1/2}/(\sin\alpha)$ of $f(x)$ -- corresponding to the
greatest pulsewidth -- occurs for the cone of observers at $x = a/b$:
\begin{equation}
    W_{\mathrm{max}} 
    = 
    2\arccos\left\{\left[\, 1 - (\sin\rho)^2/(\sin\alpha)^2 \,\right]^{1/2} \right\}
    \label{eq:A9}
\end{equation}
at $\cos\zeta = (\cos\alpha)/(\cos\rho)$.

(2) For $\alpha = \rho$, one can visualize the emission cone as
rolling around the spin axis, which is then the inner visibility
limit.  In this case, $f(x) = (\cot\alpha)[(1-x)/(1+x)]^{1/2}$; also
$x_+ = +1$ (observer looking along the spin axis) so $f(x_+) = 0$,
corresponding to $W(x_+) = \pi$.  So $W(x)$ increases monotonically
from $0$ at the outer visibility limit to $\pi$ on the spin axis in
this special case.

(3) For $\rho > \alpha$, the beam will be `always on' (unpulsed) to
observers within $(\rho - \alpha)$ of the spin axis.  So the pulse
visibility limit $x_+$ then corresponds to $W = 2\pi$, with $f(x)$
ranging from $+1$ at $x_-$ to $-1$ at $x_+$, with no minimum: $W(x)$
increases monotonically from $0$ at the outer visibility limit to
$2\pi$ at the inner one.

\subsection{Evaluating the averaging integral}

Integrating \eqn{eq:A1} for $W(\alpha,\rho,\zeta)$ over a random distribution
of viewing angles, using the observer distribution function
$(\sin\zeta)/(2f_b)$, gives the mean angular pulsewidth, for pulsars
of given $\alpha$ and $\rho$:
\begin{eqnarray}
    \left<{W}\right>\!(\alpha,\rho) 
    & = & 
    (2f_b)^{-1} \int_{|\alpha - \rho|}^{\alpha + \rho} 
    W(\rho,\alpha,\zeta)\sin\zeta \,\mathrm{d}\zeta
    \label{eq:A10} \\
    & = &
    f_b{^{-1}} \int_{x_-}^{x_+} \arccos\!\left[f(x)\right]\,\mathrm{d}x \;.
    \label{eq:A11}
\end{eqnarray}
It is integration of the observer distribution function
$(\sin\zeta)/2$ over the same range that yields the beaming factor
$f_b = \sin\alpha\,\sin\rho$.

Note for use below that
\begin{eqnarray}
    (\sin\alpha) \left[1-f^2(x)\right]^{1/2} 
    & = & 
    G^{1/2}(x)/(1-x^2)^{1/2} 
    \;\mbox{,}
    \label{eq:A12} \\
    \mbox{with~~}
    G(x) & \LRdefined & (x-x_-)(x_+-x) 
    \;\mbox{.}
    \label{eq:A13}
\end{eqnarray}
The function $G(x)$ describes an inverted parabola: it is zero at the
limits of integration $x_\pm$ and positive in between, with a maximum
value $(1-a^2)(1-b^2) = \sin^2\alpha\,\sin^2\rho \;(= f_b{^2})$ at $x
= ab = \cos\alpha\,\cos\rho$.

Applying integration by parts in \eqn{eq:A11} for
$\left<{W}\right>$ reduces the problem to one of evaluating
an integral of an algebraic function. Using $\mathrm{d}({\arccos}X) =
-(1-X^2)^{-1/2}\mathrm{d}X$ together with \eqns{eq:A4} and 
\eqNo{eq:A12} for
$\mathrm{d}f/\mathrm{d}x$ and $(1-f^2)^{1/2}$ gives
\begin{eqnarray}
    \left<{W}\right>\!(\alpha,\rho) f_{b}
    & = & x_+\arccos\!\left[f(x_+)\right]
    \nonumber \\
    && {}+{}\int_{x_-}^{x_+} \frac{x(bx-a)\;\mathrm{d}x}{(1-x^2)G^{1/2}(x)} 
    \;\mbox{,}
    \label{eq:A14}
\end{eqnarray}
because the integrated part $x\arccos\!\left[f(x)\right]$ -- which by \eqn{eq:A2} is just
$x\,W(x)/2$ -- always vanishes at the limit $x_-$, where $f=+1$ always.
The first term on the RHS of \eqn{eq:A14} is 
\begin{equation}
    \begin{array}{rccl}
        0 & \mbox{for} & \alpha > \rho & [f(x_+) = +1]  \mbox{~,}\\
        \pi/2 & \mbox{for} & \alpha = \rho & [f(x_+) = 0\;\mbox{,~}x_+ = +1]  \mbox{~,}\\
        \pi\cos(\rho - \alpha) & \mbox{for} & \alpha < \rho & [f(x_+) = -1] \mbox{~.}
    \end{array}
    \label{eq:A15}
\end{equation}

Re-arranging the integrand in \eqn{eq:A14} gives 
\begin{eqnarray}
    \lefteqn{
    \left<{W}\right>\!(\alpha,\rho) f_{b} = \frac{x_+}{2}W(x_+) 
    }&&\nonumber\\
    && {}+{} \int_{x_-}^{x_+} \left(\frac{b-a}{2(1-x)} + \frac{b+a}{2(1+x)} -b \right) 
    \frac{\mathrm{d}x}{G^{1/2}(x)} 
    \;\mbox{.}
    \label{eq:A16}
\end{eqnarray}

This result shows that we can employ the following indefinite
integrals, in which $\Delta$ and $G$ are defined by \eqns{eq:A7} and
\eqNo{eq:A13}:
\begin{equation}
    \int G^{-1/2}(x)\,\mathrm{d}x 
    = 
    \arcsin\!\left[ (x-ab)/\Delta \right]
    {\mbox{~and}} 
    \label{eq:A17} 
\end{equation}
\begin{eqnarray}
    \lefteqn{	
    \int \frac{\mathrm{d}x}{(1 \pm x)G^{1/2}(x)} 
    = 
    \int \frac{\mathrm{d}x}{(1 \pm x)\left[ (x-x_-)(x_+-x) \right]^{1/2}}
    }\nonumber \\
    & = &
    \frac{2}{|\,a \pm b\,|} \arctan \sqrt \frac{(x - x_-)(1 \pm x_+)}{(x_+ - x)(1 \pm x_-)} 
    \;\;\mbox{,}
    \label{eq:A18}
\end{eqnarray}
where $x_\pm = ab \pm \Delta$ and \eqn{eq:A7} for $\Delta$
have been used.

So the required definite integrals are
\begin{eqnarray}
    \int_{x_-}^{x_+}G^{-1/2}(x)\,\mathrm{d}x 
    & = &
    \pi 
    \mbox{~,~~and}
    \label{eq:A19} \\
    \int_{x_-}^{x_+} \frac{\mathrm{d}x}{(1 \pm x)G^{1/2}(x)} 
    & = &
    \frac{\pi}{\,|a \pm b\,|} 
    \;\mbox{.}
    \label{eq:A20}
\end{eqnarray}
Thanks to these definite integrals, \eqn{eq:A16} for the
observer-averaged pulsewidth evaluates to
\begin{equation}
    \left<{W}\right>\!(\alpha,\rho) f_{b}
    = 
    \frac{x_+}{2}W(x_+) + \frac{\pi}{2}[\, {\mathrm{sign}}(b-a) + 1 - 2b \,] 
    \;\mbox{.}
    \label{eq:A21}
\end{equation}
That is:
\begin{eqnarray}
    \lefteqn{
       \left<{W}\right>\!(\alpha,\rho) f_{b} =
       }&&\nonumber\\
    &&
    \left\{
    \begin{array}{rll}
        \left[\, 1 - \cos\rho \,\right]\pi & \mbox{for} & \alpha\geq\rho 
	\;\mbox{,}\\
        \left[\, \cos(\rho - \alpha) - \cos\rho \,\right]\pi & \mbox{for} & \alpha\leq\rho
	\;\mbox{,}\\
    \end{array}
    \right.
    \label{eq:A22}
\end{eqnarray}
as used in the main text -- \eqn{eq:PWint} above.

\label{lastpage}
\end{document}